\documentstyle [preprint,epsf,aps,eqsecnum,floats]{revtex}

\makeatletter
\def\footnotesize{\@setsize\footnotesize{10.0pt}\xpt\@xpt
\abovedisplayskip 10\p@ plus2\p@ minus5\p@
\belowdisplayskip \abovedisplayskip
\abovedisplayshortskip  \z@ plus3\p@
\belowdisplayshortskip  6\p@ plus3\p@ minus3\p@
\def\@listi{\leftmargin\leftmargini
\topsep 6\p@ plus2\p@ minus2\p@\parsep 3\p@ plus2\p@ minus\p@
\itemsep \parsep}}
\long\def\@makefntext#1{\parindent 5pt\hsize\columnwidth\parskip0pt\relax
\def\strut{\vrule width0pt height0pt depth1.75pt\relax}%
$\m@th^{\@thefnmark}$#1}
\long\def\@makecaption#1#2{%
\setbox\@testboxa\hbox{\outertabfalse %
\reset@font\footnotesize\rm#1\penalty10000\hskip.5em plus.2em\ignorespaces#2}%
\setbox\@testboxb\vbox{\hsize\@capwidth
\ifdim\wd\@testboxa<\hsize %
\hbox to\hsize{\hfil\box\@testboxa\hfil}%
\else %
\footnotesize
\parindent \ifpreprintsty 1.5em \else 1em \fi
\unhbox\@testboxa\par
\fi
}%
\box\@testboxb
} %
\global\firstfigfalse
\global\firsttabfalse
\def\tabular{\let\@halignto\@empty\@tabular}
\def\endtabular{\crcr\egroup\egroup $\egroup}
\expandafter \def\csname tabular*\endcsname #1{\def\@halignto{to#1}\@tabular}
\expandafter \let \csname endtabular*\endcsname = \endtabular
\def\@tabular{\leavevmode \hbox \bgroup $\let\@acol\@tabacol
   \let\@classz\@tabclassz
   \let\@classiv\@tabclassiv \let\\\@tabularcr\@tabarray}
\def\endtable{%
\global\tableonfalse\global\outertabfalse
{\let\protect\relax\small\vskip2pt\@tablenotes\par}\xdef\@tablenotes{}%
\egroup
}%

\def\da{d_{\rm A}}
\def\ga{g_{\rm A}}
\def\ca{C_{\rm A}}
\def\MSbar{$\overline{\rm MS}$}
\def\gammaE{{\gamma_{\rm\scriptscriptstyle E}}}
\def\LE{{\cal L}_{\rm E}}
\def\eps{\epsilon}
\def\bball{{\rm ball}}
\def\sun{{\rm sun}}
\def\hard{{\rm hard}}
\def\qcd{{\rm qcd}}
\def\resum{{\rm resum}}
\def\bare{{\rm bare}}
\def\l{{\rm L}}
\def\t{{\rm T}}
\def\sqed{{\rm sqed}}
\def\PiT{\Pi^{(T)}}
\def\PiZ{\Pi^{(0)}}
\def\lnmub{\ln{\bar\mu\over4\pi T}}
\def\alphas{\alpha_{\rm s}}
\def\alphaw{\alpha_{\rm w}}
\def\intd3q{
   \mu^{2\epsilon}\!\!\int {d^{3-2\epsilon}q \over (2 \pi)^{3-2\epsilon}}
}

\def\sumint{\hbox{$\sum$}\!\!\!\!\!\!\int}

\begin {document}

\preprint {UW/PT-94-03}

\title{The three-loop free energy for pure gauge QCD}
\author{Peter Arnold and Chengxing Zhai}

\address{ Department of Physics, FM-15,
    University of Washington,
    Seattle, Washington 98195
    }%
\maketitle

\date {July 1994}

\begin{abstract}

We compute the free energy density for pure non-Abelian gauge theory
at high temperature and zero chemical potential.  The three-loop result
to $O(g^4)$ is
\begin{eqnarray}
    F &=& \da T^4 {\pi^2\over9} \Biggr \{
        {-}{1 \over 5}
	{+}\left ({\ga \over 4 \pi} \right )^2
	{-}{16 \over \sqrt{3}} \left ({\ga \over 4 \pi} \right )^3
	{-}48 \left ({\ga \over 4 \pi} \right )^4
              \ln\left(\ga\over 2\pi\sqrt{3}\right)
\nonumber\\
	&& \qquad {+}\left ({\ga \over 4 \pi} \right )^4 \Biggr [
	{22 \over 3} \lnmub
	{+}{38 \over 3} {\zeta'(-3) \over \zeta (-3)}
	{-}{148 \over 3} {\zeta'(-1) \over \zeta (-1)}
	{-}4 \gammaE
	{+}{64 \over 5} \Biggr ]
	{+}O(\ga^5) \Biggr \} \,,
\nonumber
\end{eqnarray}
where $T$ is the temperature,
$\zeta$ is the Riemann zeta function,
$g_A \equiv g(\bar\mu) \ca^{1/2}$,
$\bar\mu$ is the \MSbar\ renormalization scale,
$g(\bar\mu)$ is the corresponding coupling constant,
and $\da$ and $\ca$ are the dimension and Casimir
of the adjoint representation.
We examine the sensitivity of this result to the choice of
renormalization scale $\bar\mu$.
We also give a result for the free energy of scalar $\phi^4$ theory,
correcting a result previously given in the literature.
\end{abstract}

\newpage
\section{Introduction}

\begin {figure}
\vbox
    {%
    \begin {center}
	\leavevmode
	
	\epsfbox [150 250 500 550] {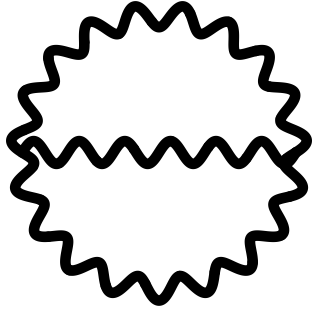}
    \end {center}
    \caption
	{%
	\label {figa}
        A 2-loop contribution to the free energy.
	}%
    }%
\end {figure}

The perturbative expansion of the free energy of hot non-Abelian gauge theory
is of the form
\begin {equation}
  F \sim T^4 [ c_0 + c_2 g^2 + c_3 g^3 + (c'_4 \ln g + c_4) g^4
          + O(g^5) ] \,,
\label {Fform}
\end {equation}
where the $c_i$ are numerical coefficients (with some dependence on the
choice of renormalization scale).  The leading term is just the free
energy of an ideal, ultrarelativistic gas of bosons.  The first effect of
interactions appears at $O(g^2)$ and can be computed from two-loop diagrams
such as fig.~\ref{figa}.  To compute to higher order requires reorganizing
perturbation theory to account for Debye screening of electric fields in the
plasma and yields terms non-analytic in $g^2$ such as $O(g^3)$ and $O(g^4 \ln
g)$.  The full $O(g^4)$ term requires a 3-loop calculation, and a full
accounting of Debye screening at 3 loops would produce the $O(g^5)$
terms.  And that's it; perturbation theory is believed incapable of pushing
the calculation to any higher order.  Beginning with four loops, infrared
problems associated with magnetic confinement appear and
{\it non}-perturbative $O(g^6)$ contribution to the free energy.%
\footnote{
  For a review of this, and also of the previously mentioned reorganization of
  perturbation theory due to Debye screening, see sec. IV of ref.~\cite{gpy}
  and also ref.~\cite{Kapusta}.
}
A complete
three-loop calculation of the free energy therefore has the special
significance that it's the best anyone will ever do with perturbation theory.
In this paper, our goal is slightly more modest.  We shall only tackle the
$O(g^4)$ contribution from three loops and leave the $O(g^5)$ contribution for
another day.

Another interest of the three-loop calculation is that $O(g^4)$ is the
first order that begins to implement the renormalization-scale independence
of the free energy.  The coupling in (\ref{Fform}) is really $g(\mu)$
where $\mu$ is some renormalization scale, and some of the coefficients
depend on $\ln\mu$.  The leading term that depends on the interaction
is order $g^2(\mu)$, and by itself depends logarithmically on our choice
of $\mu$.  A change in this term due to a small change in renormalization
scale,
\begin {equation}
     g^2(\mu') = g^2(\mu) + \beta_0 g^4(\mu) \ln{\mu'^2\over\mu^2} + \cdots
     \,,
\end {equation}
is compensated by changes in higher-order contributions, first starting
at $O(g^4)$.  The $O(g^4)$ result should therefore have a flatter dependence
on $\mu$ than the $O(g^2)$ result.  By checking this claim, we can get some
idea of the theoretical uncertainties of lower-order calculations and perhaps
learn some qualitative lessons that will carry over to other thermal
quantities.

The $O(g^3)$ piece of the free energy of non-Abelian gauge theory was
previously obtained by Kapusta \cite{Kapusta g3},
and the $O(g^4 \ln g)$ piece by Toimela \cite{Toimela}.
We shall compute an analytic result for the full $O(g^4)$
contribution.  In somewhat related work, Corian\`o and Parwani
\cite{Coriano&Parwani}
have recently studied high-temperature QED and numerically
extracted the $O(g^4)$ contribution, and Parwani \cite{Parwani e5}
has also found the
$O(g^5)$ piece.  (Unlike in non-Abelian gauge theory, the
perturbation series in QED does not break down after $g^5$.)
We shall only study pure gauge theory in this paper
and do not include any fermions.  Fermions will be included in a later
work.

In the next section, we warm up to our task by computing the
$O(g^4)$ contribution to the free energy in pure scalar theory.
The result for the basic, three-loop scalar diagram will be essential
to the later gauge theory calculation, and we shall step through our
technique for evaluating it analytically.  We shall also briefly
review the reorganization of the perturbation theory to account for
the scalar analog of the Debye mass.
In section \ref{nonabelian section},
we turn to non-abelian gauge theory and show how many
3-loop diagrams can be reduced to the scalar case.
We then discuss how to evaluate the exceptions, which are
two-particle-reducible diagrams.  Finally, in section \ref{discussion section}
we discuss our
results and examine the renormalization scale dependence.
The details of several calculations needed along the way are relegated to
appendices.

Throughout this paper we shall find it convenient to work almost
exclusively in the Euclidean (imaginary time) formulation of thermal
field theory.  We shall conventionally refer to four-momenta with
capital letters $K$ and to their components with lower-case letters:
$K = (k_0,\vec k)$.  Unless explicitly noted otherwise, all four-momenta
are Euclidean with discrete frequencies $k_0 = 2 \pi n T$.


\section {Scalar Theory}

\subsection {Basics}

Consider the theory of a real-scalar field with Euclidean Lagrangian
\begin {equation}
   \LE = {1\over2} (\partial\phi)^2
       + {1\over4!} g^2 \phi^4 \,,
\end {equation}
where we consider temperatures large enough that any zero-temperature
mass can be neglected.  The $O(g^4)$ contribution to the free energy
of this theory has been computed numerically by Frenkel, Saa, and
Taylor \cite{Frenkel}.%
\footnote{
  Note that our $g^2$ is 4!\ times their $g^2$ and that our $\eps$ is half of
  theirs.
}
In this section, we show how to obtain the result
analytically and also correct an error in the derivation of
Frenkel {\it et al.}

\begin {figure}
\vbox
    {%
    \begin {center}
	\leavevmode
	
	\epsfbox [150 250 500 500] {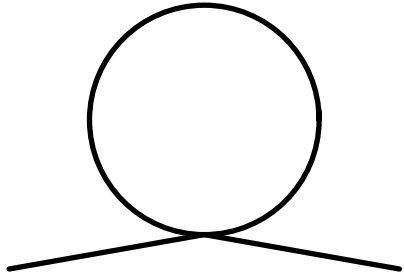}
    \end {center}
    \caption
	{%
	\label {figb}
        1-loop contribution to the scalar thermal mass.
	}%
    }%
\end {figure}

At high temperature, the scalar picks up a thermal contribution
${1\over24} g^2 T^2$
to its effective mass from the one-loop diagram of fig.~\ref{figb}.
It is inefficient to do perturbation theory with zero-temperature scalar
propagators, which do not account for this effect.  We follow
refs.~\cite{Frenkel,Parwani}\ and rewrite the Lagrangian as%
\footnote{
  For a short review in a slightly different context, that also contrasts
  this resummation scheme with the slightly different one we shall use in
  the next section for gauge theories, see ref.~\cite{Arnold&Espinosa}.
}
\begin {eqnarray}
   \LE &=& {\cal L}_0
         + {1\over4!} g^2 \phi^4
         - {1\over48} g^2 T^2 \phi^2 \,,
\\
   {\cal L}_0 &=& {1\over2} \left[
         (\partial\phi)^2 + {1\over24} g^2 T^2 \phi^2 \right] \,,
\end {eqnarray}
where the thermal mass has simply been added in and subtracted out so that
nothing is changed.  Now treat ${\cal L}_0$ as the {\it unperturbed}
Lagrangian and the last term as a perturbation.  This reorganization of
the perturbative expansion is necessary to get a well-behaved expansion
in $g$.
One can imagine including yet-higher order corrections to the thermal
mass in ${\cal L}_0$ above, but this is unnecessary and we shall not do
so.

We regularize the theory by working in $d=4-2\eps$ dimensions with
the modified minimal subtraction (\MSbar) scheme.  This corresponds to doing
minimal subtraction (MS) and then changing the MS scale $\mu$ to the
\MSbar\ scale $\bar\mu$ by the substitution
\begin {equation}
   \mu^2 = {e^{\gammaE} \bar\mu^2 \over 4\pi} \,.
\end {equation}
In dimensional regularization,
the one-loop thermal mass generated by fig.~\ref{figb} is
\begin {equation}
   m^2 = {1\over2} g^2 \sumint_P {1\over P^2} \,,
\label {m2 defn}
\end {equation}
where the integral-summation sign above is shorthand for the Euclidean
integration
\begin {equation}
   \sumint_P \to
   \mu^{2\eps} T \sum_{p_0} \int {d^{3-2\eps} p \over (2\pi)^{3-2\eps}}
\end {equation}
and the sum is over $p_0 = 2\pi n T$ for all integers $n$.  Our
reorganized Lagrangian is
\begin {eqnarray}
   \LE &=& {\cal L}_0
         + {1\over2} (Z_1-1) (\partial\phi)^2
         + {1\over4!} \mu^{2\eps} Z_2 g^2 \phi^4
         - {1\over2} m^2 \phi^2 \,,
\label {scalar resummation}
\\
   {\cal L}_0 &=& {1\over2} \left[
         (\partial\phi)^2 + m^2 \phi^2 \right] \,,
\end {eqnarray}
where $Z_1$ and $Z_2$ are the usual zero-temperature multiplicative
renormalizations:
\begin {eqnarray}
  Z_1 &=& 1 + O(g^4) \,,
\\
  Z_2 &=& 1 + {3\over2\eps} {g^2\over(4\pi)^2} + O(g^4) \,.
\end {eqnarray}

\begin {figure}
\vbox
    {%
    \begin {center}
	\leavevmode
	
	\epsfbox [150 200 500 550] {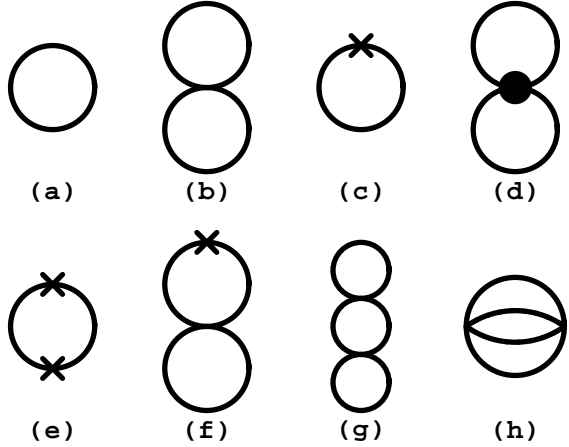}
    \end {center}
    \caption
	{%
	\label {figc}
        Diagrams contributing to the free energy in scalar $\phi^4$ theory.
        Dots represent zero-temperature counter-terms, and crosses represent
        the ``thermal counter-term'' arising from the last term of
        (\protect\ref{scalar resummation}).
	}%
    }%
\end {figure}

The diagrams contributing to the free energy $F$ through three loops are shown
in fig.~\ref{figc},
where all propagators represent the reorganized propagators of
${\cal L}_0$.  The sum of these diagrams give $-F$.
All of the diagrams except the last, the basketball diagram,
are simple because they factorize into one-loop integrals.  Diagrams
(e-g) are particularly simple because they cancel each other
at $O(g^4)$.
Diagram (a) represents the contribution to $-F$ of a non-interacting
gas of bosons of mass $m$, and its high-temperature expansion is
well-known \cite{Dolan&Jackiw}:
\begin {eqnarray}
   -{1\over2} \sumint_P \ln(P^2+m^2)
     = {\pi^2\over90} T^4 - {1\over24} m^2 T^2 && + {1\over12\pi} m^3 T
       + {1\over64\pi^2} \left[ {1\over\eps}
                                + 2\lnmub + 2 \gammaE \right] m^4
\nonumber\\ &&
       + O(m^6/T^2, \eps) \,.
\label {free F expanded}
\end {eqnarray}
The one-loop integral needed for the remaining diagrams is obtained%
\footnote{
  For a sketch of an alternative derivation directly in Euclidean space,
  see for example section III.D of ref.~\cite{Arnold&Espinosa}.
}
by differentiation with respect to $m^2$:
\begin {eqnarray}
   \sumint_P {1\over P^2+m^2}
     = {1\over12} (1+\eps \iota_\eps) T^2
       &&
       - {1\over4\pi} m T
       - {1\over(4\pi)^2} m^2 \left[ {1\over\eps}
           + 2\lnmub + 2\gammaE \right]
\nonumber\\
       + O(m^4/T^2, \eps m T, \eps^2 T^2) \,.
\label {quad expanded}
\end {eqnarray}
We have shown only those terms that can contribute to the free energy at
$O(g^4)$, but this requires introducing a term of $O(\eps)$ that
wasn't needed in (\ref{free F expanded}).  The coefficient $\iota_\eps$
of this term is less well known, and it is worth taking a moment to
focus on the $m=0$ case and review its simple derivation:
\begin {eqnarray}
   \sumint_P {1\over P^2} &=&
   \mu^{2\eps} \int {d^{3-2\eps}p\over(2\pi)^{3-2\eps}} {1\over p^2}
      + 2 \mu^{2\eps} T \sum_{n=1}^\infty
           \int {d^{3-2\eps}p\over(2\pi)^{3-2\eps}} {1\over (2\pi n T)^2 + p^2}
\nonumber\\
    &=& 2 \mu^{2\eps} T
           \int {d^{3-2\eps}p\over(2\pi)^{3-2\eps}} {1\over (2\pi T)^2 + p^2}
           \sum_{n=1}^\infty n^{1-2\eps}
\nonumber\\
    &=& {T^2 \over 2 \sqrt{\pi}} \left(\mu^2\over\pi T^2\right)^\eps
          \Gamma\left( - {\textstyle{1\over2}} + \eps \right)
          \zeta(-1+2\eps)
\nonumber\\
    &=& {T^2\over12} \left[ 1 + \eps \left(
          2\lnmub + 2 {\zeta'(-1) \over \zeta(-1)} + 2
        \right) \right] + O(\eps^2) \,,
\label {quad result}
\end{eqnarray}
so that
\begin {equation}
    \iota_\eps =
          2\lnmub + 2 {\zeta'(-1) \over \zeta(-1)} + 2 \,.
\end {equation}
With these tools, all of the diagrams but the last are straightforward.
The individual contributions of each diagram are summarized in
appendix~\ref{graphs appendix}.  Frenkel {\it et al.}\ \cite{Frenkel}
did not properly account for the $\iota_\eps$ term of
(\ref{quad expanded}) when evaluating diagrams (a--c).%
\footnote{
   More specifically, because their definition of $m^2 = {1\over24} g^2 T^2$
   differs from (\ref{m2 defn}) at $O(\eps)$,
   their thermal counterterm
   does not exactly cancel the one-loop diagram of fig.~\ref{figb} and they
   should have an extra term in their eq.~(14).  If one instead uses our
   definition (\ref{m2 defn}), then their eq.~(14) is correct but there
   should be an extra term in the one-loop pressure in their eq.~(11).
}

\subsection {The basketball diagram}
\label{bball section}

The last diagram of fig.~\ref{figc} is more difficult.
One simplification occurs
because the diagram remains infrared convergent when the mass $m$ in the
propagators is set to zero.  This means that the presence of $m$ has only
a sub-leading effect on the diagram, since $m$ itself is $O(g)$.  We can
ignore $m$ here if we are interested in the free energy only to $O(g^4)$,
and the basketball diagram is then proportional to
\begin {equation}
   I_\bball \equiv \sumint_{PQK} {1\over P^2 Q^2 K^2 (P+Q+K)^2} \,.
\label {Ibball def}
\end {equation}

Our attack on this integral starts with the observation that
the basketball diagram of fig.~\ref{figc}(h) requires three independent
four-momentum integrations if evaluated in momentum space but only
{\it one} four-space integration if instead evaluated
in configuration space.  This suggests the diagram may be more tractable
in configuration space.  Indeed it would be if the diagram were
ultraviolet (UV) convergent and we could set $\eps$ to zero.
The configuration space propagator in four dimensions, with period
$1/T$ in Euclidean time, is the relatively simple function
\begin {equation}
   \Delta(\tau,\vec r) = {T\over4\pi r} \,
     {\sinh(2\pi r T) \over [\cosh(2\pi rT) - \cos(2\pi \tau T)]} \,,
\end {equation}
but in $4{-}2\eps$ dimensions it is a nightmare.  We therefore
need to first subtract out the UV divergent pieces, and evaluate them
separately, so that we can then evaluate the remainder in four dimensions.
We found, however, that making these subtractions is more convenient in
momentum space than in configuration space.  As a result, our derivation
mixes the use of momentum and configuration space.  First,
we shall always
treat the Euclidean time direction in frequency space.  In configuration
space for the remaining, spatial dimensions, the propagator $1/P^2$
then has a very simple form in exactly four dimensions:
\begin {equation}
   \Delta(p_0,\vec r) = {e^{-|p_0|r} \over 4 \pi r} \,.
\label {3d prop}
\end {equation}
Our approach will be to start with the the momentum space form
(\ref{Ibball def}) of the basketball integral, convert it step by step
into a configuration space form, and make needed subtractions as we
go along.

\subsubsection {A careless derivation}

To simplify the presentation, let's first forget about
the UV subtractions and run through the derivation pretending that
it makes sense in exactly four dimensions despite the UV divergences.
We'll later step through it a second time, handling the divergences
more carefully.  We start by noting that, by a shift of variables,
the momentum integral (\ref{Ibball def}) can be written in the form
\begin {equation}
   I_\bball = \sumint_P [\Pi(P)]^2 \,,
\label {bball pipi form}
\end {equation}
where
\begin {equation}
   \Pi(P) \equiv \sumint_Q {1\over Q^2 (P+Q)^2} \,.
   \label {pi def}
\end {equation}
Now let's evaluate $\Pi(P)$ using configuration space and (\ref{3d prop}):
\begin {eqnarray}
   \Pi(P) &=& T \sum_{q_0} \int d^3 r \, e^{i \vec p \cdot \vec r}
                \Delta(q_0,\vec r) \, \Delta(p_0+q_0,\vec r)
\nonumber\\
   &=& {T\over(4 \pi)^2} \sum_{q_0} \int d^3 r \, {1\over r^2}
           e^{i \vec p \cdot \vec r} e^{-|q_0|r} e^{-|p_0+q_0|r}
\nonumber\\
   &=& {T\over(4 \pi)^2} \int d^3 r \, {1\over r^2}
           e^{i \vec p \cdot \vec r} (\coth \bar r + |\bar p_0|) e^{-|p_0|r}
   \,,
\label {careless pi}
\end {eqnarray}
where we have introduced the dimensionless variables
\begin {equation}
   \bar r \equiv 2 \pi T r \,,
   \qquad
   \bar p_0 \equiv p_0 / 2 \pi T \,.
\end {equation}
Plugging this into (\ref{bball pipi form}), the $\vec p$ integral becomes
trivial, producing
\begin {eqnarray}
   I_\bball &=& {T^3 \over (4\pi)^4} \sum_{p_0} \int d^3 r \, r^{-4}
               (\coth \bar r + |\bar p_0|)^2 e^{-2 |p_0| r}
\nonumber\\
   &=& {T^4 \over 32\pi^2} \int\nolimits_0^\infty d\bar r \,
            \bar r^{-2} \left(
                \coth^2 \bar r - \coth\bar r \,\, \partial_{\bar r}
                + \textstyle{1\over4} \partial^2_{\bar r}\right)
            \coth \bar r \,.
\end {eqnarray}
As we shall discuss later, integrals like this can be performed analytically
when they are convergent.  The present result doesn't make any sense, however,
because the UV behavior ($\bar r \to 0$) of the integrand makes the
integral divergent.  We shall now repeat the above derivation while
making necessary subtractions as we go along.

\subsubsection{Subtraction of UV divergences}

Let's start with the expression (\ref{careless pi}) for $\Pi(P)$.
This integral is logarithmically divergent in the ultraviolet.
As usual with one-loop integrals at finite temperature, however, it
can be made finite simply by subtracting out the zero-temperature
contribution.  So we write
\begin {equation}
   \Pi(P) = \PiZ(P) + \PiT(P) \,,
\end {equation}
where $\PiZ(P)$ is the zero-temperature result
\begin {equation}
   \PiZ(P) = \mu^{2\eps} \int {d^dQ\over(2\pi)^d} {1\over Q^2(P+Q)^2}
   = {1\over(4\pi)^2} \left( 4\pi\mu^2\over P^2 \right)^\eps
        \left( {1\over\eps} + 2 - \gammaE + O(\eps) \right)
     \,.
\end {equation}
In four dimensions, $\PiT(P)$
can be obtained from (\ref{careless pi}) by subtracting out its
$T \to 0$ limit (with $P$ fixed):
\begin {equation}
   \PiT(P) =
     {T\over(4 \pi)^2} \int d^3 r \, {1\over r^2}
           e^{i \vec p \cdot \vec r} \left(\coth \bar r - {1\over\bar r}\right)
           e^{-|p_0|r} + O(\eps)\,.
\label {PiT integral}
\end {equation}

We now split our computation of the basketball diagram into
\begin {equation}
   I_\bball = \sumint [\PiT]^2 + 2 \sumint \PiT \PiZ
              + \sumint [\PiZ]^2 \,.
\label {bball split}
\end {equation}
Though $\PiT$ is finite in four dimensions, the first term above
is not because $\PiT \sim 1/P^2$ as
$P{\to}\infty$; the first term therefore has a logarithmic UV
divergence.  The large $P$ ({\it i.e.}\ $P \gg T$) behavior of
$\PiT$ is easy to extract by staring at the definition (\ref{pi def})
of $\Pi$.  The dominant contribution comes from routing the large
momentum $P$ solely through one of the two propagators and then
integrating over the relatively small momentum $Q \alt T$ in the
other propagator:
\begin {equation}
   \PiT(P) \to {2\over P^2} \left(\sumint {1\over Q^2}\right)^{(T)}
            =  {2\over P^2}       \sumint {1\over Q^2} \,.
\label {large p limit}
\end {equation}
A more rigorous derivation may be found in appendix~\ref{pi appendix}.
This limit
is not restricted to four dimensions, so we are now in a position to
subtract out the UV divergence in our integral:
\begin {equation}
   \sumint' [\PiT]^2
   = \sumint'_P \Biggl\{ [\PiT(P)]^2
        - \left({2\over P^2}\sumint_Q{1\over Q^2}\right)^2 \Biggr\}
   + \sumint'_P \left({2\over P^2}\sumint_Q{1\over Q^2}\right)^2 \,.
\label {bballTT split}
\end {equation}
The first integral is now UV finite, and so we might hope to evaluate it in
exactly four dimensions.  However, it is not also {\it infrared} finite
if we evaluate
the $p_0 = 0$ term of the frequency sum;
for $p_0=0$, the subtraction we made
diverges linearly with $\vec p$ in the infrared.
We shall therefore treat the
$p_0=0$ mode separately and put primes on integrals, as we have
above, to denote that this mode is excluded:
\begin {equation}
   \sumint'_P \to
   \mu^{2\eps} T \sum_{p_0 \not= 0} \int {d^{3-2\eps} p \over (2\pi)^{3-2\eps}}
   \,.
\end {equation}
We can now evaluate the first term of (\ref{bballTT split})
in exactly four dimensions.

The leading large $P$ behavior (\ref{large p limit})
of $\PiT$ is related to the leading
small $r$ behavior of the integrand in (\ref{PiT integral}) and is
given by
\begin {equation}
   \PiT(P) \to
     {T\over(4 \pi)^2} \int d^3 r \, {1\over r^2} \,
           e^{i \vec p \cdot \vec r} \,\, {\bar r\over 3} \,\,
           e^{-|p_0|r} + O(\eps) \,.
\label {PiT limit integral}
\end {equation}
Following the same steps as in the careless derivation of the
previous section, we then obtain
\begin {eqnarray}
   \sumint'_P && \left\{ [\PiT(P)]^2
        - \left({2\over P^2}\sumint_Q{1\over Q^2}\right)^2 \right\}
\nonumber\\
   && \qquad\qquad
   = {T^3 \over (4\pi)^4} \sum_{p_0 \not= 0} \int d^3 r \, r^{-4}
               \left[ \left(\coth \bar r - {1\over\bar r}\right)^2
                          - \left(\bar r\over 3\right)^2 \right]
               e^{-2 |p_0| r}
     + O(\eps)
\nonumber\\
   && \qquad\qquad
   = {T^4 \over 32\pi^2} \int\nolimits_0^\infty d\bar r \, \bar r^{-2}
               \left[ \left(\coth \bar r - {1\over\bar r}\right)^2
                          - \left(\bar r\over 3\right)^2 \right]
               (\coth \bar r - 1)
     + O(\eps) \,.
\end {eqnarray}
This integral is both IR and UV convergent and can be evaluated using
the techniques of appendix~\ref{integration appendix} to give
\begin {eqnarray}
   \sumint'_P && \left\{ [\PiT(P)]^2
        - \left({2\over P^2}\sumint_Q{1\over Q^2}\right)^2 \right\}
\nonumber\\ && \qquad
   = {1\over(4\pi)^2} \left(T^2\over12\right)^2 \left[
        - 16 {\zeta'(-3)\over\zeta(-3)}
        + 1152 \zeta'(-2)
        + 24 {\zeta'(-1)\over\zeta(-1)}
        - 8\gammaE + {28\over 15}
   \right]
   + O(\eps) \,.
\label {bballTT part a}
\end {eqnarray}
The last term in (\ref{bballTT split}) is easily evaluated in $4{-}2\eps$
dimensions using (\ref{quad result}) and
\begin {equation}
   \sumint_P {1\over P^4}
    = {1\over(4\pi)^2} \left[ {1\over\eps} +
          2\lnmub + 2\gammaE
       \right] + O(\eps) \,,
\label {quart result}
\end {equation}
which may be obtained in a manner similar to (\ref{quad result}).

Completing the derivation of the $[\PiT]^2$ contribution
to $I_\bball$ now just requires
adding in the contribution of the $p_0=0$ mode, which is UV convergent and
does not require any subtractions:
\begin {eqnarray}
   T \int {d^{d-1}p \over (2\pi)^{d-1}} [\PiT(0,p)]^2
   &=& {T^4\over32\pi^2} \int\nolimits_0^\infty d\bar r \, \bar r^{-2}
      \left( \coth \bar r - {1\over\bar r} \right)^2
   + O(\eps)
\nonumber\\
   &=&  {1\over(4\pi)^2} \left(T^2\over12\right)^2 \left[
        - 1152 \zeta'(-2) \right]
   + O(\eps) \,,
\label {bballTT part b}
\end {eqnarray}
where we have again used the techniques of appendix~\ref{integration appendix}
to do the integral.
Putting together (\ref{quad result}, \ref{bballTT split}, \ref{bballTT part a},
\ref{quart result}, \ref{bballTT part b}) then gives
\begin {equation}
   \sumint [\PiT]^2
   = {1\over(4\pi)^2} \left(T^2\over12\right)^2 \left[
        {4\over\eps} + 24 \lnmub
        - 16 {\zeta'(-3)\over\zeta(-3)}
        + 40 {\zeta'(-1)\over\zeta(-1)}
        + {268\over15} \right]
   + O(\eps) \,.
\label {PiT2 integral}
\end {equation}

Now that we've covered the basic ideas of our technique, we'll leave the
evaluation of the remaining two terms in (\ref{bball split}) to
appendix~\ref{bball appendix}.  The final result for the basketball integral
(\ref{Ibball def}) is
\begin {equation}
   I_\bball
   = {1\over(4\pi)^2} \left(T^2\over12\right)^2 \left[
        {6\over\eps} + 36 \lnmub
        - 12 {\zeta'(-3)\over\zeta(-3)}
        + 48 {\zeta'(-1)\over\zeta(-1)}
        + {182\over5} \right]
   + O(\eps) \,.
\label{bball result}
\end {equation}
This agrees with the numerical result of ref.~\cite{Frenkel}.  We should
mention that our analytic result can also be obtained
from the integrals generated by a real-time analysis, such as in
ref.~\cite{Frenkel}, and we show how to do this in
appendix~\ref{real time appendix}.
We have found it simpler to stick to Euclidean space, however,
to evaluate diagrams involving double poles $1/P^4$ which will
appear later in gauge theories.

\subsection {The result}

Putting together all the diagrams, which are independently tabulated
in appendix~\ref{graphs appendix},
one finds that the free energy in $\phi^4$ theory
at high temperature is
\begin{eqnarray}
    F &=& T^4 {\pi^2\over9} \Biggr \{
        {-}{1\over10}
	{+}{1\over8} \left( g \over 4 \pi \right)^2
	{-}{1 \over \sqrt{6}} \left( g \over 4 \pi \right)^3
\nonumber\\
	&& \qquad {+}\left( g \over 4 \pi \right)^4 \Biggr [
	{-}{3 \over 8} \lnmub
	{+}{1 \over 4} {\zeta'(-3) \over \zeta (-3)}
	{-}{1 \over 2} {\zeta'(-1) \over \zeta (-1)}
	{-}{1 \over 8} \gammaE
	{+}{59 \over 120} \Biggr ]
	{+}O(g^5) \Biggr \} \,.
\end{eqnarray}

\begin {figure}
\vbox
    {%
    \begin {center}
	\leavevmode
	
	\epsfbox [150 250 500 550] {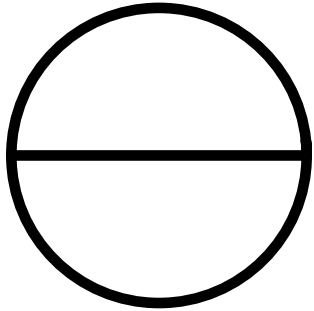}
    \end {center}
    \caption
	{%
	\label {figd}
        The setting sun diagram.
	}%
    }%
\end {figure}

Before leaving scalar theory, we should mention one other basic scalar
integral which appears in the literature
\cite{Parwani,Arnold&Espinosa} and which will be needed when we
analyze gauge theories.  It is the integral corresponding to the
setting sun diagram of fig.~\ref{figd},
\begin {equation}
   I_\sun(m_1,m_2,m_3) \equiv
   \sumint_{PQ} {1\over(P^2+m_1^2)(Q^2+m_2^2)[(P+Q)^2+m_3^2]} \,,
\label {sunset_def}
\end {equation}
evaluated to leading order in the masses.
It has previously only been evaluated
numerically \cite{Parwani}, but the same techniques we applied to the
basketball diagram can be used to obtain an analytic result.
We give the derivation in appendix~\ref{sunset appendix}, with the result that
\begin {equation}
   I_\sun = {T^2\over(4\pi)^2} \left[
              {1\over4\eps}
              + \ln\left( \bar\mu \over m_1+m_2+m_3 \right)
              + {1\over2}
            \right] + O(m,\eps) \,.
\label {Isun result}
\end {equation}


\section {Nonabelian gauge theory}
\label {nonabelian section}

We now turn to pure non-Abelian gauge theory, given by the Lagrangian
\begin {equation}
   \LE = {1\over4}
         \left( \partial_\mu A_\nu^a - \partial_\nu A_\mu^a
                + g f^{abc} A_\mu^b A_\nu^c \right)^2
         + (\hbox{gauge fixing}) \,.
\end {equation}
We shall work exclusively in Feynman gauge.
(It would be nice to explicitly verify that our results are independent of
gauge choice, but we have not done so.)
Let $\da$ and $\ca$ be the dimension and Casimir of the adjoint
representation, with $\ca$ given by
\begin {equation}
   f^{abc}f^{dbc} = \ca \delta^{ad} \,.
\end {equation}
For SU($N$), they are
\begin {equation}
   \da = N^2 - 1 \,,
   \qquad \qquad \qquad
   \ca = N       \,.
\end {equation}
It is also convenient to define the effective coupling $\ga$ of the
adjoint representation by
\begin {equation}
   \ga^2 \equiv g^2 \ca \,.
\end {equation}
As before, we shall regulate the theory with dimensional regularization in
the \MSbar\ scheme.

\begin {figure}
\vbox
    {%
    \begin {center}
	\leavevmode
	
	\epsfbox [150 350 500 500] {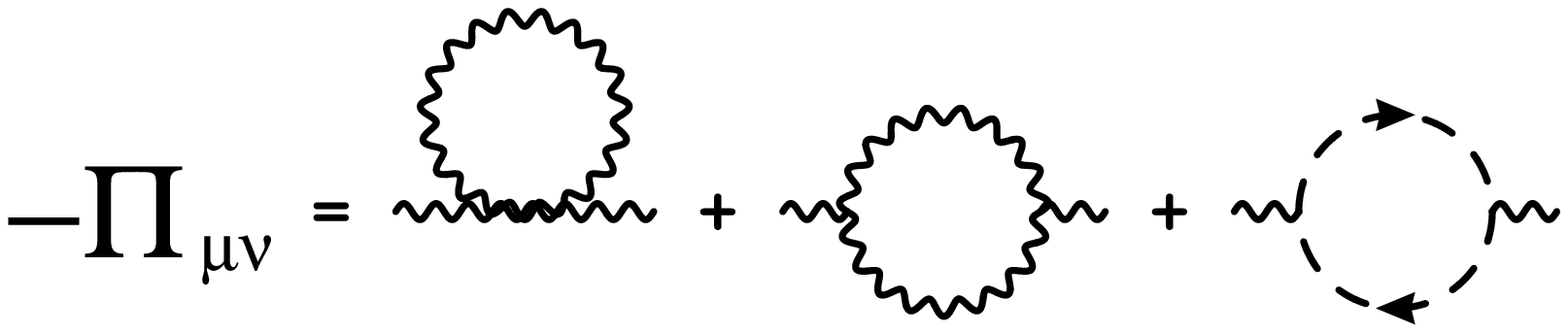}
    \end {center}
    \caption
	{%
	\label {fige}
        The one-loop gluon self energy.
	}%
    }%
\end {figure}

As in the scalar case, one-loop effects induce a thermal mass contribution.
This mass is given by the one-loop self-energy $\Pi_{\mu\nu}$ at zero
momentum, and in Euclidean space a mass is generated for $A_0$ but not
for $\vec A$.%
\footnote{
   Throughout this article, $\Pi_{\mu\nu}(0)$ will refer to the Euclidean
   limit $(p_0{=}0, \vec p{\to}0)$ and never to the limit
   $(p_0{\to}0, \vec p=0)$ which may be achieved by analytic continuation
   and which gives the mass gap for propagating plasma waves.
}
This mass $M$ is the Debye screening mass for static electric fields and may
be evaluated from the diagrams of fig.~\ref{fige} as
\begin {equation}
   M^2 \delta^{ab}
       = \Pi^{ab}_{\mu\mu}(0)
       = \ga^2 (d-2)^2 \sumint_Q {1\over Q^2} \delta^{ab} \,.
\label{M def}
\end {equation}
In four dimensions, $M^2$ is simply $\ga^2 T^2/3$.
In the gauge theory calculation, we find it calculationally convenient to
use a slightly different reorganization of perturbation theory than we did
for in the scalar case.  The success of the reorganization only depends
on the behavior of the propagator in the infrared ($p_0{=}0$, $p{\ll}T$),
where the mass cannot be treated as a perturbation.
We follow ref.~\cite{Arnold&Espinosa} and only
introduce the mass for the $p_0=0$ mode.  That is, we rewrite our Lagrangian
density, in frequency space, as
\begin {equation}
  \LE = \left(\LE + {\textstyle{1\over2}} M^2 A^a_0 A^a_0
                               \delta_{p_0^{}}\right)
                  - {\textstyle{1\over2}} M^2 A^a_0 A^a_0
                               \delta_{p_0^{}} \,,
\label {gauge resummation}
\end {equation}
where $\delta_{p_0^{}}$ is shorthand for the the Kronecker delta function
$\delta_{p_0^{},0}$.
Then we absorb the first $A_0^2$ term into our
unperturbed Lagrangian ${\cal L}_0$
and treat the second $A_0^2$ term as a perturbation.%
\footnote{
   This reorganization only helps in the evaluation of static quantities
   such as the free energy.  To evaluate time-dependent correlations in
   real time, one would need the resummation scheme of Braaten and
   Pisarski \protect\cite{Braaten&Pisarski}.
}

Since the necessity of resummation is an infrared phenomena, associated
with the mass scale $gT$, it is worth noting that the prescription
(\ref{gauge resummation}) can be naturally expressed in the language
of decoupling.  First imagine integrating out all the physics associated
with scales $\agt T$.  In particular, integrating out all of the
$p_0 \not = 0$ modes in Euclidean space generates an effective three
dimensional theory of the remaining $p_0=0$ modes.  This effective
theory will have the thermal mass for $A_0$ and other interactions induced
by the heavy modes, which can be computed to any desired order in
perturbation theory.  Only then does one finally integrate out the
$p_0=0$ modes after deciding on a sensible partition of the effective
three-dimensional Lagrangian into an unperturbed piece, containing the
thermal mass terms, and a perturbative piece.  Rather than carry out the
reduction to this effective theory explicitly, however, we find it simplest
to just introduce the reorganization (\ref{gauge resummation}).
We refer the reader to sections III.D and VI of ref.~\cite{Arnold&Espinosa}
for details of how
to implement this form of reorganization on two-loop graphs.%
\footnote{
   The reader should beware that many of the specific formulas of
   ref.~\cite{Arnold&Espinosa} are particular to Landau gauge, whereas in
   the present work we are working in Feynman gauge.
}

\begin {figure}
\vbox
    {%
    \begin {center}
	\leavevmode
	
	\epsfbox [100 100 500 700] {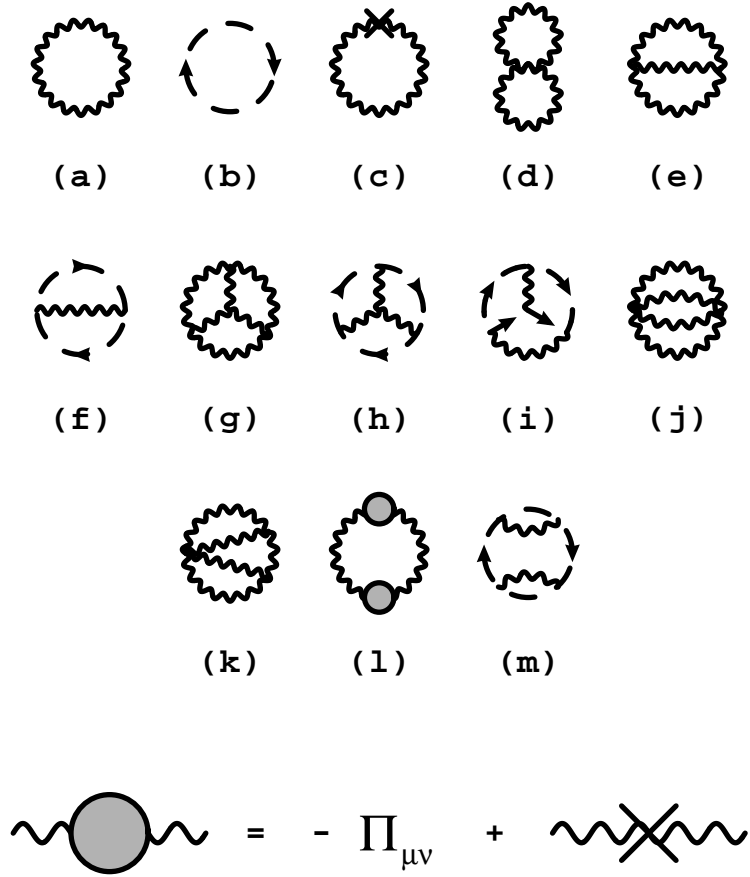}
    \end {center}
    \caption
	{%
	\label {figf}
        Diagrams contributing to the free energy in gauge theory.
        The crosses are the ``thermal counter-terms'' arising from the
        last term of (\protect\ref{gauge resummation}).
        We have not explicitly shown any zero-temperature counter-terms, and
        each diagram should be multiplied by the appropriate
        multiplicative renormalizations for vertices and propagators.
	}%
    }%
\end {figure}

\begin {figure}
\vbox
    {%
    \begin {center}
	\leavevmode
	
	\epsfbox [150 160 500 600] {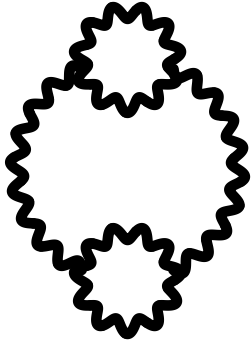}
    \end {center}
    \caption
	{%
	\label {figg}
        A diagram with infrared problems.
	}%
    }%
\end {figure}

The diagrams that contribute to the free energy are shown in fig.~\ref{figf}.
The diagrams involving only one-loop integrations are trivial, and the
resummation and the two-loop graphs can be handled by the methods of
ref.~\cite{Arnold&Espinosa}.
Let's therefore focus on the three-loop
diagrams.  The first potential problem is that some of the
individual diagrams, such as fig.~\ref{figg}, are infrared
divergent because of the masslessness of the $\vec A$ propagator.
However, the particular combination we have shown in fig.~\ref{figf}(l) is
well-behaved in the infrared since the shaded blobs,
\begin {equation}
   \Delta\Pi_{\mu\nu} \equiv \Pi_{\mu\nu}(P) - \Pi_{\mu\nu}(0) \delta_{p_0^{}}
   \,,
\end {equation}
are $O(pT)$ for $(p_0{=}0, \vec p{\to}0)$.
We can then also drop the Debye mass $M$ in evaluating the three-loop
graphs since, as in the scalar case, the corrections to the free energy
will be beyond $O(g^4)$.

All of the three loop diagrams except (l) can be reduced to the scalar
basketball integral of (\ref{Ibball def}).  For example, diagram (i) is
equal to
\begin{equation}
    -\mu^{2\eps}F^{\rm i} = - {1 \over 8} \da \ga^2 \sumint_{PQK}
	{P \cdot (Q-K) \, (P-K) \cdot Q
	\over P^2 Q^2 K^2 (P-Q)^2 (Q-K)^2 (K-P)^2} \,.
\label {reduction example}
\end{equation}
This may be reduced by (1) expanding numerator factors in terms of
denominator factors to cancel factors between numerator and denominator,
such as
\begin {equation}
    P\cdot(Q-K) = {\textstyle{1\over2}} [(K-P)^2 - K^2 - (P-Q)^2 + Q^2] \,;
\label{(i) expansion}
\end {equation}
(2) performing appropriate changes of variables to collect similar terms;
and (3) using the identity
\begin{equation}
   \sumint_P {P_{\mu} \over (P+Q)^2 (P+K)^2}
	= - {Q_{\mu} + K_{\mu} \over 2}
	\sumint_P {1 \over (P+Q)^2 (P+K)^2} \,,
\label{trick}
\end{equation}
which follows by averaging the left-hand side with itself, after
applying the change of variables $P \to {-}P{-}Q{-}K$.
Appendix~\ref{sample reduction appendix}
steps through this reduction for the example (\ref{reduction example}),
and the reductions of diagrams (g--k,m) are all tabulated in
appendix~\ref{graphs appendix}.

Unfortunately, diagram (l) cannot be reduced to the scalar basketball.
If one tries the above tricks, one finds a term of the form
\begin {equation}
    I_\hard \equiv \sumint_{PQK}
	{(Q \cdot K)^2 \over P^4 Q^2 K^2 (P+Q)^2 (P+K)^2}
\label {Ihard defn}
\end{equation}
for which the tricks fail to remove the numerator factor.
So we have a new basic integral that we must
evaluate, like the basketball integral of scalar theory.
We have found it more tractable, however, to apply our integration method
directly to the original diagram (l) because the
orthogonality of the one-loop self-energy $\Pi_{\mu\nu}(P)$
to $P_\mu$ leads to useful algebraic simplifications.
Diagram (l) is proportional to
\begin {equation}
   \da \ga^4 I_\qcd \equiv
   \sumint_P {1\over P^4} {\rm tr} [\Delta\Pi_{\mu\nu}(P)]^2 \,.
\end {equation}
The evaluation of $I_\qcd$ is somewhat similar to that of the basketball
integral $I_\bball$ and is presented in appendix~\ref{qcd appendix}.
We should mention,
however, that the derivation is more complicated and involves
a miraculous cancelation between two complicated integrals that we don't
know how to calculate individually.  The appearance and cancelation of
such complications suggests that we may still be missing the most elegant
method for making these calculations.

All the results for individual graphs are collected in
appendix~\ref{graphs appendix}, with
the final result that
\begin{eqnarray}
    F &=& \da T^4 {\pi^2\over9} \Biggr \{
        {-}{1 \over 5}
	{+}\left ({\ga \over 4 \pi} \right )^2
	{-}{16 \over \sqrt{3}} \left ({\ga \over 4 \pi} \right )^3
	{-}48 \left ({\ga \over 4 \pi} \right )^4
              \ln\left(\ga\over 2\pi\sqrt{3}\right)
\nonumber\\
	&& \qquad {+}\left ({\ga \over 4 \pi} \right )^4 \Biggr [
	{22 \over 3} \lnmub
	{+}{38 \over 3} {\zeta'(-3) \over \zeta (-3)}
	{-}{148 \over 3} {\zeta'(-1) \over \zeta (-1)}
	{-}4 \gammaE
	{+}{64 \over 5} \Biggr ]
	{+}O(\ga^5) \Biggr \} \,.
\label{F result}
\end{eqnarray}
For those who prefer $\zeta$ functions with positive arguments,
\begin {equation}
   {\zeta'(-n)\over\zeta(-n)} = \ln(2\pi) + \gammaE
       - \sum_{k=1}^n {1\over k}
       - {\zeta'(1+n)\over\zeta(1+n)} \,,
   \qquad \hbox{$n$ odd} \,.
\end {equation}


\begin {figure}
\vbox
    {%
    \begin {center}
	\leavevmode
	
	\epsfbox [150 250 500 550] {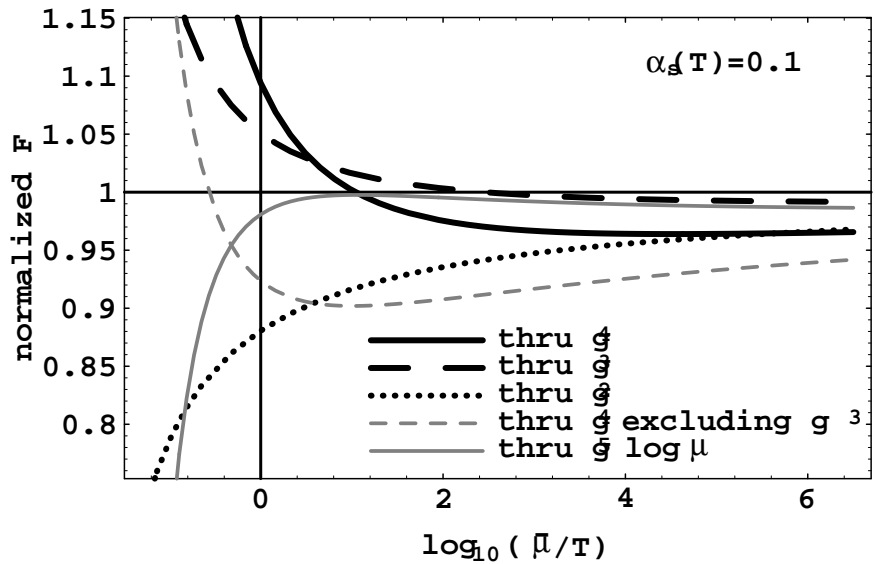}
    \end {center}
    \caption
	{%
	\label {figh1}
        The dependence of the free energy $F$ on the choice of renormalization
        scale $\bar\mu$ for pure gauge QCD with $\alphas(T) = 0.1$.
        The free energy is normalized in units of the ideal gas result
        $-\da \pi^2 T^4/45$.
        The thick solid, dashed, and dotted lines are the results for $F$
        including terms through $g^4$, $g^3$, and $g^2$ respectively.
        The light solid curve is the $g^4$ result plus the
        $g^5 \ln(\bar\mu/T)$ term required by renormalization group
        invariance.  The light dashed curve is the $g^4$ result minus the
        $g^3$ term.
	}%
    }%
\end {figure}

\section {Discussion}
\label{discussion section}

Evaluated numerically, our result (\ref{F result}) is
\begin {eqnarray}
    F = - \da {\pi^2 T^4\over 45} && \biggl[
       1
       - 0.31250 \left(\ga\over\pi\right)^2
       + 0.72168 \left(\ga\over\pi\right)^3
\nonumber
\\ && \qquad
       + \left(\ga\over\pi\right)^4 \left(
             0.93750 \ln{\ga\over\pi} - 0.14323 \ln{\bar\mu\over T}
             + 0.74582  \right)
       + O(\ga^5) \biggr] \,.
\label {F numeric}
\end {eqnarray}
We have chosen the expansion parameter $\ga/\pi$ simply because it makes
all the coefficients $O(1)$.

Now we can ask whether perturbation theory is behaving well for
physically-realized values of the couplings.  In particular, we can
investigate (1) the size of corrections from different orders for a
fixed choice of renormalization scale, such as $\bar\mu=T$, and
(2) whether higher-order results are less sensitive to the choice
of the renormalization scale $\bar\mu$ than lower-order results.
This information is summarized in fig.~\ref{figh1} for pure gauge
QCD with $\alphas(T)=0.1$, which for real QCD would correspond to
a temperature around the electroweak scale.  We have used the
two-loop renormalization group to compute $g(\mu)$:
\begin {equation}
   {1\over\ga^2(\mu)} \approx
      {1\over\ga^2(T)} - \beta_0 \ln{\bar\mu\over T}
      + {\beta_1\over\beta_0}
         \ln\left(1 - \beta_0 \ga^2(T) \ln{\bar\mu\over T}\right)
      \,,
\end {equation}
where
\begin{equation}
   \beta_0 = - {22\over 3 (4 \pi)^2} \,,
   \qquad
   \beta_1 = - {68\over 3 (4 \pi)^4} \,.
\end {equation}

At $\bar\mu=T$, the
terms of (\ref{F numeric}) for $\alphas(T){=}0.1$ are
\begin {equation}
   F = - \da {\pi^2 T^4\over 45} [
          1 - 0.12 + 0.17 + ( -0.07 + 0 + 0.11 ) + O(\ga^5)
       ] \,.
\end {equation}
The behavior of the perturbative expansion doesn't look particularly good,
though a partial cancelation between the $g^4 \ln g$ and $g^4$ terms
makes the total $O(g^4)$ contribution relatively small.
Alternatively, examine the sensitivity of the result to the
choice of renormalization scale by examining the slopes of the curves
in fig.~\ref{figh1} at $\bar\mu=T$.  Contrary to one's expectation for
a well-behaved perturbative expansion, the $O(g^4)$ result is
{\it more} sensitive to $\bar\mu$ than the $O(g^2)$ or $O(g^3)$
results.

\begin {figure}
\vbox
    {%
    \begin {center}
	\leavevmode
	
	\epsfbox [150 250 500 550] {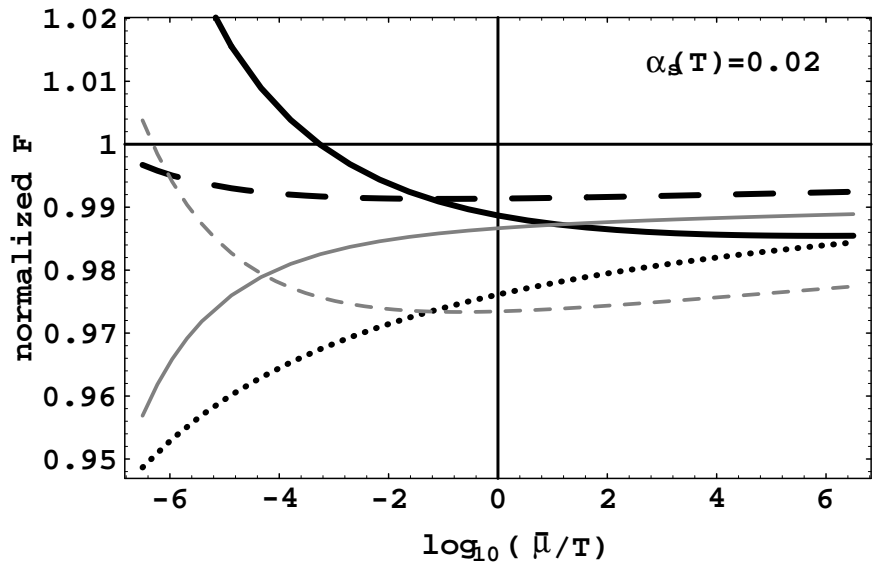}
    \end {center}
    \caption
	{%
	\label {figh2}
        The same as fig.~\protect\ref{figh1} but for $\alphas(T) = 0.02$
        [which is identical to SU(2) gauge theory with
        $\alphaw(T) \approx 1/33$].
	}%
    }%
\end {figure}

\begin {figure}
\vbox
    {%
    \begin {center}
	\leavevmode
	
	\epsfbox [150 250 500 550] {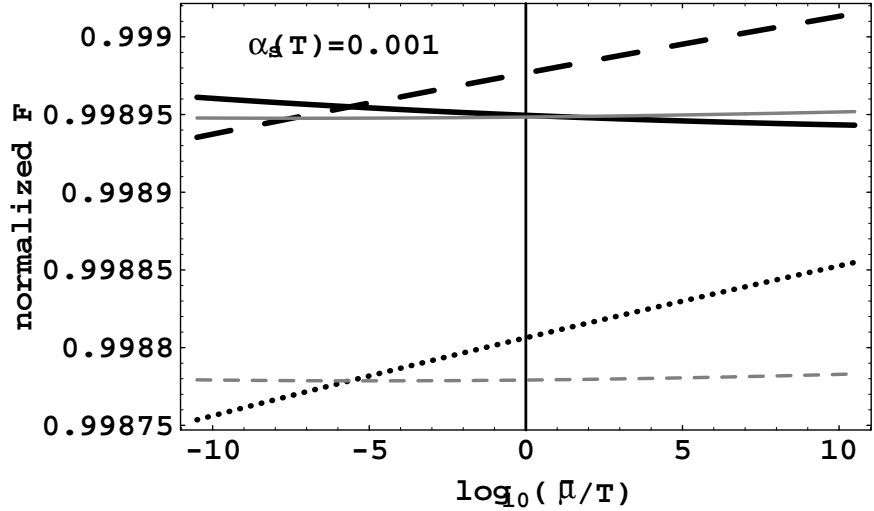}
    \end {center}
    \caption
	{%
	\label {figh3}
        The same as fig.~\protect\ref{figh1} but for $\alphas(T) = 0.001$.
	}%
    }%
\end {figure}

The $O(g^4)$ result has to be less sensitive to $\bar\mu$ if $g$ is
sufficiently small.  Fig.~\ref{figh2} shows the dependence for
$\alphas(T)=0.02$.  This is equivalent to a system of interest---pure
electroweak theory at the electroweak scale, with $\alphaw \approx
1/33$.
Yet still the $O(g^4)$ result is no less sensitive than the $O(g^2)$ result.
Fig.~\ref{figh3} shows $\alphas(T)=0.001$, where we finally see the
expected behavior.  (In six-flavor QCD, this $\alphas$ would
correspond to a temperature of $10^{387}$ GeV.)

The source of the sensitivity problem can be
found by remembering that the $g^3$ term probes different physics than
the $g^2$ term: the $g^2$ term is produced by particles with hard
thermal momenta of order $T$, while the $g^3$ term arises from the
interactions of particles with softer momentum of order $gT$.
Since the $g^3$ term is the {\it leading}-order contribution of
the physics of scale $gT$, it perhaps should be treated independently
of the $g^2$ term when discussing whether the perturbation expansion
is well-behaved.  The $g^4$ terms reduce the sensitivity of the
$g^2$ term to $\bar\mu$, and the $g^5$ contribution will be needed
to reduce the sensitivity of the $g^3$ term.  The light dashed line
in figs.~\ref{figh1}--\ref{figh2} show the result of the free energy
through $O(g^4)$ if the $g^3$ term is artificially {\it excluded}.
The sensitivity to $\bar\mu$, compared to the $O(g^2)$ result,
is indeed much better than before.

In order to put the $O(g^3)$ term back in, we have tried adding the
$g^5 \ln\mu$ term that's determined by renormalization group invariance.
The light solid lines in figs.~\ref{figh1}--\ref{figh2} represent our
result (\ref{F result}) with the addition
\begin {equation}
    \Delta F =
    \da T^4 {\pi^2\over9} \Biggr \{
        {-}{16 \over \sqrt{3}} \left(\ga \over 4\pi\right)^5
            \left[ 11 \lnmub \right]
    \Biggl\} \,.
\end {equation}
The results are much better behaved than the $O(g^4)$ results discussed
earlier.
Of course, the constant under the log at $O(g^5)$ is unknown and
will change the curves somewhat.
Our understanding of whether perturbative results are indeed well-behaved
in high-temperature QCD, for realistic coupling constants, would therefore
benefit by a true calculation to $O(g^5)$---the last order accessible to
perturbation theory.


\bigskip

This work was supported by the U.S. Department of Energy,
grant DE-FG06-91ER40614.  We thank Larry Yaffe, and
most especially Lowell Brown, for their repeated suggestions that we
try analyzing our basic diagrams in configuration space using the simple
form of the propagators in three spatial dimensions.
We also thank Rajesh Parwani and Claudio Corian\`o for useful discussions.


\newpage
\appendix

\section {Results for individual graphs}
\label {graphs appendix}
\subsection {Scalar theory}
The diagrams of fig.~\ref{figc} are given by
\begin {mathletters}%
\begin {eqnarray}
&&
   -\mu^{2\eps}F^{\rm a} =
      -{1\over2} \sumint \ln(P^2+m^2)
   \,,
\\ &&
   -\mu^{2\eps}F^{\rm b} =
      -{1\over8} g^2
      \left( \sumint {1\over P^2+m^2} \right)^2
   \,,
\\ &&
   -\mu^{2\eps}F^{\rm c} =
      {1\over2} m^2 \sumint {1\over P^2+m^2}
   \,,
\\ &&
   -\mu^{2\eps}F^{\rm d} =
      -{1\over8} \left(3 g^4\over 32\pi^2\eps\right)
      \left( \sumint {1\over P^2+m^2} \right)^2
   + O(g^6) \,,
\\ &&
   -\mu^{2\eps}(F^{\rm e} + F^{\rm f} + F^{\rm g}) =
   O(g^5) \,,
\\ &&
   -\mu^{2\eps}F^{\rm h} =
     {1\over48} g^4 I_\bball
   + O(g^5) \,.
\setcounter{eqletter}{8}    
\end {eqnarray}
\end {mathletters}

\subsection {Gauge theory}

Writing $F = \mu^{-2\eps} \da {\cal F}$ and ignoring terms of $O(\eps)$, the
diagrams of fig.~\ref{figf} are given by
\begin {mathletters}%
\label {gauge diagrams}%
\begin {eqnarray}
   -{\cal F}^{\rm a} &=&
      - {d\over2} \sumint \ln P^2 + {1\over12\pi} M^3 T
   \,,
\\
   -{\cal F}^{\rm b} &=&
      \sumint \ln P^2
   \,,
\\
   -{\cal F}^{\rm c} &=&
      -{1\over8\pi} (M_1^2 + M_2^2 + M_3^2) M T
   \,,
\\
   -{\cal F}^{\rm d} &=&
      -{1\over4} \ga^2 Z_g^2 d(d-1)
          \left( \sumint{1\over P^2} \right)^2
      +{1\over8\pi} M_1^2 M T
   \,,
\\
   -{\cal F}^{\rm e} &=&
      \ga^2 Z_g^2 \left[
         {3\over4} (d-1) \left( \sumint{1\over P^2} \right)^2
         + \delta_1
         + \delta_2
       \right]
   \,,
\\
   && \delta_1 =
       - M^2 \sumint { \delta_{p_0^{}} (1-\delta_{q_0^{}})
                       \over (P^2+M^2)Q^2(P+Q)^2           }
       + \left(d-{3\over2}\right) I_\resum
       + {1\over8\pi} {M_2^2 M T \over \ga^2 Z_g^2}
   \,,
\nonumber
\\
   && \delta_2 =
       - {1\over4} {M^2 T^2 \over (4\pi)^2}
       - M^2 \sumint{ \delta_{p_0^{}} \delta_{q_0^{}} \over
                      (P^2+M^2)(Q^2+M^2)(P+Q)^2 }
   \,,
\nonumber
\\
   -{\cal F}^{\rm f} &=&
      \ga^2 Z_g^2 \left[
         - {1\over4} \left( \sumint{1\over P^2} \right)^2
         - {1\over2} I_\resum
      \right]
      + {1\over8\pi} M_3^2 M T
   \,,
\\
   -{\cal F}^{\rm g} &=&
      \left( {5\over8} d - {23\over32} \right)
      \ga^4 I_\bball
      + O(g^5)
   \,,
\\
   -{\cal F}^{\rm h} &=&
      - {1\over16} \ga^4 I_\bball
      + O(g^5)
   \,,
\\
   -{\cal F}^{\rm i} &=&
      - {1\over32} \ga^4 I_\bball
      + O(g^5)
   \,,
\\
   -{\cal F}^{\rm j} &=&
      {3\over16} d(d-1) \ga^4 I_\bball
      + O(g^5)
   \,,
\\
   -{\cal F}^{\rm k} &=&
      - {27\over16} (d-1) \ga^4 I_\bball
      + O(g^5)
   \,,
\\
   -{\cal F}^{\rm l} &=&
      {1\over4} \ga^4 I_\qcd
      + O(g^5)
   \,,
\\
   -{\cal F}^{\rm m} &=&
      - {1\over8} \ga^4 I_\bball
      + O(g^5)
   \,.
\end {eqnarray}
\end {mathletters}
The multiplicative renormalization constant used for the coupling is
given by
\begin {equation}
    g_\bare = Z_g g
     = \left[ 1 - {11\over6} {\ga^2 \over (4\pi)^2\eps} + O(g^4) \right] g
    \,.
\end {equation}
Wave function renormalization constants are unnecessary because they cancel
between vertices and propagators.  [To this end, the most convenient choice
of $M$ is
\begin {equation}
   M^2 = \ga^2 Z_g^2 Z_A^2 (d-2)^2 \sumint{1\over P^2} \,
\end {equation}
which differs from (\ref{M def}) by the introduction of $Z_g^2$ and
the photon wave function renormalization $Z_A^2$.
At $O(g^4)$, however, this is not an issue---the factor of
$Z_g^2 Z_A^2$ in $M^2$ can be ignored for all of our diagrams.]
$M_1^2$, $M_2^2$, and $M_3^2$ denote the three pieces of $M^2$
originating, respectively, from the three diagrams in fig.~\ref{fige}.
We will not give explicit formulas for these pieces because they explicitly
cancel between diagrams (c--f).

The integrals needed above are given
by (\ref{free F expanded}, \ref{quad expanded}, \ref{bball result},
\ref{Iresum result}, \ref{H333}, \ref{H344}, \ref{Iqcd result}).
The integral $I_\resum$ is defined by
\begin {eqnarray}
   I_\resum
   &\equiv& \sumint \left[  {\delta_{p_0^{}} \over P^2+M^2}
                     - {\delta_{p_0^{}} \over P^2}     \right]
     \left[ {q_0^2 \over Q^2 (P+Q)^2} - {q_0^2 \over Q^4} \right]
\\
   &=& - \sumint \delta_{p_0^{}} {M^2\over P^4}
     \left[ {q_0^2 \over Q^2 (P+Q)^2} - {q_0^2 \over Q^4} \right]
     + O(g^3)
\label{Iresum form}
\end {eqnarray}
and will be discussed below.

The effect of the thermal mass term in the $p_0{=}0$ gauge propagator
appears fairly simply in the last terms of
(\ref{gauge diagrams}a, d) and in (\ref{gauge diagrams}c).
The case of diagram (e) is a little more complicated.
The first term of (\ref{gauge diagrams}e) is the result when
$M$ is ignored.  $\delta_1$ is the correction to this result for
the contribution to the diagram where exactly one of the three
propagators has $p_0=0$, and $\delta_2$ is the correction for
the contribution where all three have $p_0 = 0$.  So, for instance,
\begin {equation}
   \delta_1 = \left\{ {1\over4} \sumint { \delta_{p_0^{}} (1-\delta_{q_0^{}})
                                  \over (P^2+M^2)Q^2(P+Q)^2 }  \right\}
              \left[ (P-Q)_\mu \delta_{0\nu} + 2 Q_0 \delta_{\mu\nu}
                     - (2P+Q)_\nu \delta_{\mu 0} \right]^2
   - \{ M{\to}0 \}
   \,,
\end {equation}
where the factor in brackets is the triple gauge vertex.
Using the reduction tricks described after (\ref{reduction example}),
$\delta_1$ may be reduced to
\begin {eqnarray}
   \delta_1 =
       - M^2 \sumint { \delta_{p_0^{}} (1-\delta_{q_0^{}})
                       \over (P^2+M^2)Q^2(P+Q)^2           }
       &+& \left(d-{\textstyle{3\over2}}\right)
           \sumint \left[ {\delta_{p_0^{}} \over P^2+M^2}
                        - {\delta_{p_0^{}} \over P^2} \right]
           {q_0^2 \over Q^2 (P+Q)^2}
\nonumber
\\
       &+& {1\over2} \sumint {\delta_{p_0^{}}\over (P^2+M^2)} \, {1 \over Q^2}
   \,,
\end {eqnarray}
which may be recast in the form shown in (\ref{gauge diagrams}e).
$\delta_2$, and the mass effects in (\ref{gauge diagrams}f), are
calculated similarly.

$I_\resum$ is easily evaluated by taking the form
(\ref{Iresum form}) and scaling all three-momenta by $|q_0|$:
\begin {eqnarray}
   I_\resum = - M^2 && T^2 \mu^{4\eps} \sum_{q_0} |q_0|^{2d-8}
\nonumber
\\ && \times
              \int {d^{d-1} p\over (2\pi)^{d-1}} {d^{d-1} q\over(2\pi)^{d-1}}
              {1\over p^4} \left[ {1\over (1+q^2)(1+|\vec p + \vec q|^2)}
                                - {1\over (1+q^2)^2} \right]
              + O(g^3) \,.
\end {eqnarray}
Recognizing the $q_0$ sum as giving a $\zeta$-function, and that the
integrals are finite, it is easy to now take the $\eps{\to}0$ limit:
\begin {eqnarray}
   I_\resum &=& - 2 M^2 T^2 \zeta(0)
              \int {d^3 p\over (2\pi)^3} {d^3 q\over(2\pi)^3}
              {1\over p^4} \left[ {1\over (1+q^2)(1+|\vec p + \vec q|^2)}
                                - {1\over (1+q^2)^2} \right]
              + O(g^3,\eps)
\nonumber
\\
   &=& -{1\over8} {M^2 T^2\over(4\pi)^2}
       + O(g^3,\eps) \,.
\label{Iresum result}
\end {eqnarray}


\section {Large $P$ behavior of scalar $\Pi(P)$}
\label {pi appendix}

In this appendix, we shall derive the large momentum behavior of $\Pi (P)$.
Recalling the definition~(\ref{pi def}),
\begin{equation}
    \Pi (P) = \sumint_Q {1 \over Q^2 (P + Q)^2} \,,
\label{Pi_def2}
\end{equation}
one may sum over the bosonic frequency modes by the usual contour
integral trick.%
\footnote{
   See ref.~\cite{Kapusta} for a review.
}
Subtracting out the zero-temperature piece gives
the finite-temperature part of $\Pi (P)$:
\begin{equation}
    \Pi^{(T)} (P) = - \intd3q {n (q) \over q}
	\left [
        {1 \over (q - i p_0)^2 - |{\vec q} + {\vec p}\,|^2} +
        {1 \over (q + i p_0)^2 - |{\vec q} - {\vec p}\,|^2}
        \right ] \,,
\label{PiT_expression}
\end{equation}
where we have defined the thermal bosonic factor
\begin{equation}
    n (q) \equiv {1 \over e^{q/T} - 1} \,.
\label {Bose factor}
\end{equation}
The exponential fall-off of $n (q)$ for large $q$ ensures that only
$q \alt T$ is important.
For $P \gg T$, we can then expand the denominators in~(\ref{PiT_expression}):
\begin{eqnarray}
    \Pi^{(T)} (P) &=& \intd3q {n (q) \over  q}
 	\left [{2 \over P^2}
	+ {8 (i p_0 q + {\vec p} \cdot {\vec q}\,)^2
	\over P^6} \right ] + O (T^6/P^6)
\nonumber\\
	&=& {2 \over P^2} \intd3q {n (q) \over  q}
	+ {8 \over P^6} \left ({p^2 \over d{-}1}{-}p_0^2 \right )
	\intd3q n (q) q
\nonumber\\ && \qquad\qquad
        +O (T^6/P^6) \,.
\label{Pi_exp}
\end{eqnarray}
The momentum integrals we need are of the form
\begin{eqnarray}
    J_\alpha &\equiv& T^{-3-\alpha} \intd3q n (q) q^\alpha
\nonumber\\
    &=& {  2 \over (4 \pi)^{3/2}
             \Gamma\left( {\textstyle{3\over2}} - \epsilon \right)  }
	\left ({4 \pi \mu^2 \over T^2} \right )^\epsilon
	\int_0^{\infty} dq\, q^{2-2\epsilon+\alpha}
	{1 \over e^q - 1}
\nonumber\\
	&=& \left ({4 \pi \mu^2 \over T^2} \right )^\epsilon
	{\Gamma (3{-}2 \epsilon{+}\alpha) \zeta (3{-}2 \epsilon{+}\alpha)
	\over 4 \pi^{3/2}
             \Gamma\left( {\textstyle{3\over2}} - \epsilon \right)  }
        \,,
\label{J_result}
\end{eqnarray}
and our large $P$ expansion is
\begin{equation}
    \Pi^{(T)} (P) = 2 J_{-1} {T^2 \over P^2} + 8 J_1 {T^4 \over P^6}
	\left ({p^2 \over d-1} - p_0^2 \right )
	+ O (T^6 / P^6) \,.
\label{PiT_exp_result}
\end{equation}
We note here that
\begin{equation}
    J_{-1} T^2 = \sumint_Q {1 \over Q^2} \,,
\end{equation}
and so the limit~(\ref{large p limit}) in the main text is the same as
the leading piece of~(\ref{PiT_exp_result}) above.


\section {Integrals of hyperbolic functions}
\label {integration appendix}

In this appendix, we discuss how to evaluate convergent integrals of
the form
\begin {equation}
    I = \int\nolimits_0^\infty\! dx \left(
           \sum_{m,n} c_{mn} x^m \coth^n x
           + \sum_m d_m x^m e^{-a_m x}
        \right) \,,
\end {equation}
where $a_m$, $c_{mn}$, and $d_m$ are constants.
We shall evaluate such integrals term by term.  Since the individual terms
will in general be divergent, we first regulate in a manner similar
to dimensional regularization by rewriting $I$ as
\begin {equation}
    I = \lim_{\delta\to0}
    \int\nolimits_0^\infty\! dx \left(
       \sum_{m,n} c_{mn} x^{m+\delta} \coth^n x
       + \sum_m d_m x^{m+\delta} e^{-a_m x}
    \right) \,.
\label {a hyperbolic integral}
\end {equation}
$\delta$ will be used to independently regulate both the $x{\to}0$ and
$x{\to}\infty$ pieces of the integrals, again similar to dimensional
regularization.  Now we can compute three basic regulated integrals:
\begin {equation}
   \int\nolimits_0^\infty\! dx\, x^z = 0 \,,
\end {equation}
\begin {equation}
   \int\nolimits_0^\infty\! dx\, x^z \coth x
   = \int\nolimits_0^\infty\! dx\, x^z
       \left[1 + 2\sum_{k=1}^\infty e^{-2kx} \right]
   = 2^{-z} \Gamma(z+1) \, \zeta(z+1) \,,
\end{equation}
\begin {equation}
   \int\nolimits_0^\infty\! dx\, x^z e^{-ax}
   = a^{-1-z} \Gamma(1+z) \,.
\end {equation}
The rest of the integrals we need are obtainable recursively by
\begin {eqnarray}
   \int\nolimits_0^\infty\! dx\, x^z \coth^n x
   &=& \int\nolimits_0^\infty\! dx\, x^z \left[
          -{1\over n-1} \partial_x \left( \coth^{n-1} x \right)
          + \coth^{n-2} x
       \right]
\\
   &=& \int\nolimits_0^\infty\! dx\, \left[
          {z\over n-1} x^{z-1} \coth^{n-1} x
          + x^z \coth^{n-2} x
       \right] \,.
\end {eqnarray}
After assembling the individual terms of a particular integral
(\ref{a hyperbolic integral}), it is straightforward to expand
in $\delta$ and take the limit $\delta{\to}0$.


\section {Completion of the calculation of $I_\bball$}
\label {bball appendix}

In section \ref{bball section},
the basket ball diagram was split into three terms,
\begin{equation}
    I_{\rm ball} = \sumint \left[\Pi^{(T)} \right ]^2
	+ 2 \, \sumint \Pi^{(T)} \Pi^{(0)}
	+ \sumint \left[\Pi^{(0)} \right ]^2 \,,
\label{split2}
\end{equation}
and the first term was evaluated.
Here we shall evaluate the two remaining terms.

We first calculate the second term in~(\ref{split2}). To apply
the calculational method of section \ref{bball section}, we must first
subtract the ultraviolet divergences.
$\Pi^{(0)}$ is given by
\begin{equation}
    \Pi^{(0)} = A \left ( {4 \pi \mu^2 \over P^2} \right )^\epsilon
\label{Pi0_exp}
\end{equation}
where
\begin{equation}
    A = {1 \over (4 \pi)^2} \left ({1 \over \epsilon}
	+ 2 - \gammaE \right )  +  O(\epsilon) \,.
\end{equation}
Because $\Pi^{(T)} (P) \sim 1/P^2$ at large momentum $P$,
the second term of~(\ref{split2}) is quadratically divergent and
so requires two subtractions. Using our result~(\ref{PiT_exp_result}) for
the large momentum expansion of $\Pi^{(T)}$, we rewrite this term as
\begin{equation}
    \sumint_P \Pi^{(T)} \Pi^{(0)} = I_{\rm a} + I_{\rm b} + I_{\rm c} \,,
\end{equation}
where
\begin{equation}
    I_{\rm a} \equiv \sumint_P \left [\Pi^{(0)} (P)
	- {1 \over (4 \pi)^2\epsilon} \right ]
	 \left [\Pi^{(T)} (P)
	 - \Pi_{\rm UV}^{(T)} (P) \right ] \,,
\label{I_a_def}
\end{equation}
\begin{equation}
    I_{\rm b} \equiv {1 \over (4 \pi)^2\eps} \,\,
	\sumint_P \left [\Pi^{(T)} (P)
	 - \Pi_{\rm UV}^{(T)} (P) \right ] \,,
\label{I_b_def}
\end{equation}
\begin{equation}
    I_{\rm c} \equiv \sumint_P \Pi^{(0)} (P)
	 \Pi_{\rm UV}^{(T)} (P) \,,
\label{I_c_def}
\end{equation}
and
\begin{equation}
    \Pi_{\rm UV}^{(T)} (P) \equiv
	2 J_{-1} {T^2 \over P^2} + (1 - \delta_{p_0^{}})
	8 J_1 {T^4 \over P^6} \left ({p^2 \over d-1} - p_0^2 \right ) \,.
\label{PiT_UV_def}
\end{equation}
$I_{\rm a}$ is ultraviolet and infrared finite, and so it can
be evaluated in $d=4$.
Note that we have used one less subtraction in (\ref{PiT_UV_def}) for the
$p_0=0$ mode.

Using the integral representation~(\ref{PiT integral}) for $\Pi^{(T)}$
gives
\begin{eqnarray}
    I_{\rm a} =&& {T \over (4 \pi)^4} \sumint_P
	\left[ \ln \left ({4 \pi \mu^2 \over P^2} \right )
               + 2 - \gammaE \right]
	\int d^3 r {1 \over r^2}
	e^{i {\vec p} \cdot {\vec r}} \left [
	\coth \bar r - {1 \over \bar r}
	- {\bar r \over 3} + (1-\delta_{p_0^{}})
	{\bar r^3 \over 45} \right ] e^{-|p_0| r}
\nonumber\\ && \qquad
	+ O (\epsilon) \,,
\label{I_a_exp}
\end{eqnarray}
where we have expressed the large $P$ behavior (\ref{PiT_UV_def}) of
$\Pi^{(T)} (P)$ in terms of three-dimensional coordinate space
integrals.
In fact, it is easy to see that the $P{\to}\infty$ behavior of
$\Pi^{(T)} (P)$ in~(\ref{PiT integral}) simply corresponds
to the $\bar r{\to}0$ behavior of $\coth \bar r - 1/\bar r$,
and the subtractions in~(\ref{I_a_exp}) simply reflect the small
$\bar r$ expansion
\begin{equation}
    \coth \bar r = {1 \over \bar r} + {\bar r \over 3} - {\bar r^3 \over 45}
    + \cdots \,.
\end{equation}
The $\vec p$ integral in (\ref{I_a_exp}) is trivial for the terms that don't
involve $\ln P^2$: it just gives $\delta(\vec r)$, which in turn gives
zero.
The ${\vec p}$ integral involving the logarithm can be evaluated
by first writing
\begin{eqnarray}
    \int {d^3 p \over (2 \pi)^3}
	e^{i {\vec p} \cdot {\vec r}}
	\ln {4 \pi \mu^2 \over p^2 + p_0^2}
	&=& \left .
	{d \over d \alpha} \left [ \int_0^{\infty} dp \,
	{p^2 \over 2 \pi^2} {\sin pr \over pr}
	\left ({4 \pi \mu^2 \over p^2 + p_0^2} \right )^{\alpha}
	\right ] \right |_{\alpha =0}
\nonumber\\
	&=& {1 \over 4 i \pi^2 r} \left .
	{d \over d \alpha} \left [ \int_{-\infty}^{\infty}
	dp \, p \, e^{ipr}
	\left ({4 \pi \mu^2 \over p^2 + p_0^2} \right )^{\alpha}
	\right ] \right |_{\alpha =0} \,.
\end{eqnarray}
Deforming the contour to wrap around the cut in
the upper half of the complex plane gives
\begin{eqnarray}
   \int {d^3 p \over (2 \pi)^3}
        e^{i {\vec p} \cdot {\vec r}}
	\ln {4 \pi \mu^2 \over p^2 + p_0^2}
	&=& {1 \over 2 \pi^2 r}
	\left . {d \over d \alpha} \left [\sin(\pi\alpha)
	\int_{|p_0|}^{\infty} dq \, q e^{-qr} {1 \over (q^2 - p_0^2)^{\alpha}}
	\right ] \right |_{\alpha =0}
\nonumber\\
	&=& {1 \over 2 \pi r} \left ({|p_0| \over r}
	+ {1 \over r^2} \right ) e^{-|p_0| r} \,.
\label{tricky}
\end{eqnarray}
Inserting this result into (\ref{I_a_exp}) and carrying out the $p_0$
sum yields
\begin{eqnarray}
    I_{\rm a} = {T^2 \over (4 \pi)^2} {1\over2}
        \int_0^{\infty} {d \bar r \over \bar r^3} \Biggr \{ &&
        \left (\coth \bar r{-}{1 \over \bar r} - {\bar r \over 3}
           {+}{\bar r^3 \over 45} \right )
        \left (1{-}{\bar r \over 2} {d \over d\bar r} \right )
           (\coth\bar r - 1)
\nonumber
\\ && \qquad
        + \left (\coth \bar r - {1 \over \bar r}
              {-}{\bar r \over 3} \right ) \Biggr \}
        + O(\eps) \,.
\end{eqnarray}
The integral can be computed by the method of Appendix B to give
\begin{equation}
    I_{\rm a} = {1 \over (4 \pi)^2}
        \left ({T^2 \over 12} \right )^2
        \left [
	  {8 \over 5} {\zeta'(-3) \over \zeta (-3)}
          - 4 {\zeta'(-1) \over \zeta (-1)}
	  + {12 \over 5} \gammaE
          - {46 \over 15}
        \right ]
    + O(\eps) \,.
\label{I_a_result}
\end{equation}

We now study $I_{\rm b}$ defined by~(\ref{I_b_def}).
Though the $P$ integration converges, we still need to evaluate
$I_{\rm b}$ in $d{=}4{-}2\epsilon$ dimensions because of the overall
factor of $1/\epsilon$.
Writing $\Pi^{(T)}$ as $\Pi-\Pi^{(0)}$ gives
\begin{eqnarray}
    I_{\rm b} &=& {1 \over (4 \pi)^2} {1 \over \epsilon} \,\,
        \sumint_P \left [\sumint_Q {1 \over Q^2 (Q + P)^2}
	- \Pi^{(0)} (P) - \Pi_{\rm UV}^{(T)} (P) \right ]
\nonumber\\
	&=& {T^4 \over (4 \pi)^2} {1 \over \epsilon}
	\left \{ - J_{-1}^2
	- A \, \left ({4 \pi \mu^2 \over T^2} \right )^{\epsilon}
	S_0 (\epsilon) - {8 J_1 \over d-1} \left [
	S_0 (2) - d \, S_1 (3) \right ] \right \} \,,
\label{I_b_exp}
\end{eqnarray}
where we have defined
\begin{eqnarray}
    S_n (\alpha) &\equiv& T^{2\alpha-4-2n}
	\sumint_Q {q_0^{2n} \over Q^{2\alpha}}
\label{S_n_def}\\
        &=& \left ({\mu^2 \over \pi T^2} \right )^\epsilon
	{(2 \pi)^{3+2n-2\alpha}  \over  4 \pi^{3/2} \Gamma (\alpha)}
        \Gamma\left(-{\textstyle{3\over2}}{+}\alpha{+}\epsilon\right)
	\zeta ({-}3{-}2n{+}2\alpha{+}2\epsilon)
    \,.
\label{S_n_result}
\end{eqnarray}
Taking the limit $\epsilon \to 0$,
\begin{equation}
    I_{\rm b} = {1 \over (4 \pi)^2} \left ({T^2 \over 12} \right )^2
        \left [
	{8 \over 5} {\zeta'(-3) \over \zeta (-3)}
        - 4 {\zeta'(-1) \over \zeta (-1)}
	+ {12 \over 5} \gammaE
        - {46 \over 15} \right ]
    + O(\eps) \,.
\label{I_b_result}
\end{equation}
(We do not have an explanation for the fact that $I_{\rm a} = I_{\rm b}$.)

Using the explicit expression~(\ref{Pi0_exp}) for $\Pi^{(0)}$, $I_{\rm c}$
is
\begin{eqnarray}
    I_{\rm c} &=& 2 A \, T^4 \,
	\left ({4 \pi \mu^2 \over T^2} \right )^\epsilon
	\left \{ J_{-1} S_0 (1{+}\epsilon) +
        {4 J_1 \over d{-}1} \left [S_0 (2{+}\epsilon)
	- d \, S_1 (3{+}\epsilon) \right ] \right \}
\nonumber\\
	&=&  {1 \over (4 \pi)^2}
	\left ({T^2 \over 12} \right )^2
	\left [{4 \over 5\epsilon}
	+ {24 \over 5} \ln {{\bar \mu} \over 4 \pi T}
	- {12 \over 5} {\zeta'(-3) \over \zeta(-3)}
	+ 12 {\zeta'(-1) \over \zeta(-1)}
	- {24 \over 5} \gammaE
	+ 13 \right ] + O (\epsilon) \,.
\label{I_c_result}
\end{eqnarray}
Adding the results~(\ref{I_a_result}), (\ref{I_b_result}), and
(\ref{I_c_result}) produces
\begin{equation}
    \sumint \Pi^{(T)} \Pi^{(0)}
	= {1 \over (4 \pi)^2}
	\left ({T^2 \over 12} \right )^2
	\left [{4 \over 5 \epsilon}
	+ {24 \over 5} \ln {{\bar \mu} \over 4 \pi T}
	+ {4 \over 5} {\zeta'(-3) \over \zeta(-3)}
	+ 4 {\zeta'(-1) \over \zeta(-1)}
	+ {103 \over 15} \right ]
    + O(\eps) \,.
\label{secondterm}
\end{equation}
It is straightforward to also evaluate the third term in~(\ref{split2}):
\begin{eqnarray}
    \sumint \left[\Pi^{(0)} \right ]^2
	&=& A^2 \, T^4 \,
	\left ({4 \pi \mu^2 \over T^2} \right )^{2\epsilon}
	S_0 (2 \epsilon)
\nonumber\\
	&=& {1 \over (4 \pi)^2}
	\left (T^2 \over 12 \right )^2
	\left [{2 \over 5\epsilon}
	+ {12 \over 5} \ln {{\bar \mu} \over 4 \pi T}
	+ {12 \over 5} {\zeta'(-3) \over \zeta(-3)}
	+ {24 \over 5} \right ]
   + O(\eps) \,.
\label{thirdterm}
\end{eqnarray}
Assembling~(\ref{PiT2 integral}), (\ref{secondterm}),
and (\ref{thirdterm}) gives our final result (\ref{bball result}).


\section {Real-time calculation of $I_\bball$}
\label {real time appendix}

In ref.~\cite{Frenkel}, $I_{\rm ball}$ is expressed as
\begin{equation}
    I_{\rm ball} = {1 \over (4 \pi)^2} \left(T^2 \over 12\right )
           \left [{6 \over \epsilon}
	          + 18 \ln {\pi \mu^2 \over T^2}
	          + 18 \gammaE + 6 - 36 {\zeta'(2) \over \zeta(2)}
           \right]
	+ N T^4
    + O(\eps) \,,
\label{Frenkelball}
\end{equation}
where $N$ is given by
\begin{eqnarray}
   N &=& 4 \, {\cal P}\! \int
        {d^4P\over(2\pi)^3} {d^4Q\over(2\pi)^3} {d^4K\over(2\pi)^3}
        \,\delta(P^2)\,\delta(Q^2)\,\delta(K^2)\,
        {n(p)\, n(q)\, n(k) \over (P+Q+K)^2}
\nonumber\\
   &=& {1\over32\pi^6} \int dp \,dq \,dk\, n(p)\, n(q)\, n(k)
	\Bigl[(p{+}q{+}k)\ln(p{+}q{+}k)
	-(p{+}q{-}k) \ln |p{+}q{-}k|
\nonumber\\
	&& \qquad
        -(q+k-p) \ln |q+k-p|
	-(k+p-q) \ln |k+p-q| \Bigr]
\label {N_def}
\end{eqnarray}
and was evaluated numerically to get
\begin {equation}
   N \approx {14.17 \over 32\pi^6} \,.
\end{equation}
In (\ref{N_def}), unlike the rest of this paper, $P$ refers to
Minkowski rather than Euclidean four-momentum, with metric
$P^2=-p_0^2+p^2$.
${\cal P}$ denotes that the integrals are to be performed with the
principal value prescription.
The $n(p)$ are the usual Bose factors but in units where
$T{=}1$:
\begin {equation}
   n(p) = {1 \over e^p - 1} \,.
\end {equation}
The second equality in (\ref{N_def}) is obtained by doing the trivial
$p_0$, $q_0$, and $k_0$ integrals and then doing the angular integrals.
Their final result was then obtained by numeric integration.  We shall
show how to obtain the same result analytically starting from
(\ref{N_def}).

We start by making use of the peculiar identity that $(P+Q+K)^2$ can
be replaced by $4 |\vec p + \vec q + \vec k|^2$ in the first line of
(\ref{N_def}) for {\it any} function $n(p)$ of $p=|\vec p|$ to give
\begin {equation}
   N =  8
        \int {d^3 p \over (2 \pi)^3}
	{d^3 q \over (2 \pi)^3}
	{d^3 k \over (2 \pi)^3}
	{n (p) \over 2p}
	{n (q) \over 2q}
	{n (k) \over 2k}
	{1 \over |{\vec p} + {\vec q} + {\vec k}|^2} \,.
\label{N rewritten}
\end {equation}
This identity can be proved by brute force by doing the angular integrals
in (\ref{N rewritten}) by the same steps ref.~\cite{Frenkel} used for
(\ref{N_def}) and verifying that the result is the same as (\ref{N_def}):
\begin{eqnarray}
    N &=& {1 \over \pi^2} \int {d^3 p \over (2 \pi)^3}
        \int {d^3 q \over (2 \pi)^3}
	\int_0^{\infty} dk {n(p) \over 2p}
	{n(q) \over 2q} n(k)
	{1 \over |{\vec p} + {\vec q}\,|}
	\ln {|{\vec p} + {\vec q}\,| + k
	\over ||{\vec p} + {\vec q}\,| - k|}
\nonumber\\
    &=& {1 \over 32 \pi^6} \int_0^{\infty} dp\, dq\, dk\,
	n(p)\, n(q)\, n(k) \int_{|p-q|}^{p+q} d|\vec p+\vec q|
	\ln {|\vec p+\vec q\,|+k \over ||\vec p+\vec q\,|{-}k|}
\nonumber\\
    &=& {1 \over 32 \pi^6} \int_0^{\infty} dp\, dq\, dk\,
        n(p)\, n(q)\, n(k) \Bigl[
	(p{+}q{+}k) \ln (p{+}q{+}k)
	- (p+q-k) \ln |p+q-k|
\nonumber\\
	&& \qquad \qquad
        + (|p-q|-k)\ln ||p-q|-k|
        - (|p{-}q|{+}k) \ln (|p{-}q|{+}k) \Bigr]
    \,.
\label{N_bar_result0}
\end{eqnarray}
Sadly, we do not have a more elegant derivation.

Now take (\ref{N rewritten}) and convert it to coordinate space using
the Fourier transform
\begin{equation}
    \int {d^3 p \over (2 \pi)^3}
	{n (p) \over p}
	e^{i {\vec p} \cdot {\vec r}}
    = {1 \over 2 \pi^2 r} \int_0^{\infty} dp \,\,
	{\sin pr \over e^{p} -1}
    = {1 \over 4 \pi r}
	\left ( \coth (\pi r) - {1 \over \pi r} \right ) \,.
\label{bose_FT}
\end{equation}
Eq.~(\ref{N rewritten}) then becomes
\begin{eqnarray}
    N &=& \int d^3 r \left [{1 \over 4 \pi r} \left (
	\coth \pi r - {1 \over \pi r} \right ) \right]^3
	{1 \over 4 \pi r}
\nonumber\\
	&=& {1 \over 4} {1 \over (4 \pi)^2}
	\int_0^{\infty} d x \,
	{1 \over x^2} \left (\coth x - {1 \over x}
	\right )^3
\nonumber\\
    &=& {1 \over 4} {1 \over (4 \pi)^2}
	\left [
	- {1 \over 3} {\zeta'(-3) \over \zeta(-3)}
        + {1 \over 3} {\zeta'(-1) \over \zeta(-1)}
	- {7 \over 45} \right ]
\nonumber\\
    &\approx& {14.1723\over 32\pi^6} \,.
\label {N_result}
\end{eqnarray}
Eqs.~(\ref{Frenkelball}) and (\ref{N_result}) give the same result for
the basketball integral as (\ref{bball result}), which was our result
using the Euclidean formalism.


\section {Derivation of $I_\sun$}
\label {sunset appendix}

Let us evaluate the integral for the setting sun diagram defined
by (\ref{sunset_def}) in a similar way as
the basketball integral.
As usual, we work in the limit that masses are all much smaller than
$T$, and we shall denote their order of magnitude simply as $O(m)$.
Only the leading term in the $m/T$ expansion will be calculated.
To this order, the masses can be taken to be zero except in the
$p_0{=}q_0{=}0$ contribution, where the mass cuts off a logarithmic
infrared divergence.
But it is convenient to keep only one mass non-zero in the
$p_0{=}q_0{=}0$ contribution and to set $m_2{=}m_3{=}0$.
The discrepancy introduced by doing so is easily computed
in coordinate space to be \cite{Arnold&Espinosa}
\begin{eqnarray}
    T^2 \int {d^3 p \over (2 \pi)^3} && {d^3 q \over (2 \pi)^3}
	{1 \over p^2 + m_1^2}
	\left [{1 \over (q^2+m_2^2)\left[ ({\vec p} + {\vec q}\,)^2
	+ m_3^2 \right ]}
	- {1 \over q^2 ({\vec p} + {\vec q}\,)^2} \right ]
    + O(m,\eps)
\nonumber\\
	&&= T^2 \int d^3 r {1 \over (4 \pi r)^3} e^{- m_1 r}
	\left [e^{-(m_2 + m_3) r} - 1 \right ]
    + O(m,\eps)
\nonumber\\
	&&= {T^2 \over (4 \pi)^2}
	\ln {m_1 \over m_1 + m_2 + m_3}
    + O(m,\eps) \,.
\label{error_result}
\end{eqnarray}

\subsection {Quick derivation using the contour trick}

We now need to compute
\begin {equation}
   I_\sun = \sumint {\delta_{p_0^{}} \delta_{q_0^{}} \over
            (P^2+m_1^2) (Q^2+m_2^2) [(P+Q)^2+m_3^2]}
       + \sumint {1 - \delta_{p_0^{}} \delta_{q_0^{}} \over
            P^2 Q^2 (P+Q)^2 }
       + O(m,\eps) \,.
\label {quick split 1}
\end {equation}
We start with the purely three-dimensional contribution \cite{Shappy},
\begin {equation}
   \sumint {\delta_{p_0^{}} \delta_{q_0^{}} \over
            (P^2+m_1^2) (Q^2+m_2^2) [(P+Q)^2+m_3^2]}
    = {T^2 \over (4 \pi)^2}
	\left [{1 \over 4 \epsilon}
	+ \ln {{\bar\mu} \over m_1 + m_2 + m_3}
	+ {1 \over 2} \right ] + O (\epsilon) \,,
\label {H333}
\end {equation}
which follows from
\begin {eqnarray}
   \sumint {\delta_{p_0^{}} \delta_{q_0^{}} \over
            (P^2+m_1^2) Q^2 (P+Q)^2}
   &=& \sumint {\delta_{p_0^{}} \over (P^2+m_1^2)}
       {\mu^{2\eps} T \over p^{1+2\eps}}
       { \Gamma\left(\textstyle{1\over2}+\eps\right)
         \Gamma^2\left(\textstyle{1\over2}-\eps\right)
         \over
         (4\pi)^{{3\over2}-\eps}
         \Gamma(1-2\eps)
       }
\nonumber\\
    &=& {T^2 \over (4 \pi)^2}
	\left [{1 \over 4 \epsilon}
	+ \ln {{\bar\mu} \over m_1}
	+ {1 \over 2} \right ] + O (\epsilon)
\label {H333m1}
\end{eqnarray}
and (\ref{error_result}).
For the second term of (\ref{quick split 1}), note that if dimensional
regularization is used to regulate the infrared as well as the
ultraviolet then
\begin {equation}
   \sumint \sumint {\delta_{p_0^{}} \delta_{q_0^{}} \over
            P^2 Q^2 (P+Q)^2}
   = 0
\end {equation}
simply by dimensional analysis.  (There is no scale to make up for the
$\mu^{2\eps}$.)  So
\begin {eqnarray}
   \sumint {1 - \delta_{p_0^{}} \delta_{q_0^{}} \over
               P^2 Q^2 (P+Q)^2 }
   &=& \sumint {1 \over P^2 Q^2 (P+Q)^2}
\nonumber\\
   &=&
   - {3\over2}\, \sumint {d^{d-1}p \over (2\pi)^{d-1} 2p}
               {d^{d-1}q \over (2\pi)^{d-1} 2q}
     [n(p)-n(-p)] [n(q)-n(-q)]
\nonumber\\ && \qquad \times
     \left[ {1\over |\vec p+\vec q\,|^2 - (p+q)^2}
          + {1\over |\vec p-\vec q\,|^2 - (p-q)^2} \right]
\nonumber\\ &&
   + ~ \hbox{(T independent)}
\nonumber \\
   &=& 0 \,,
\end {eqnarray}
where we have used the standard contour trick \cite{Kapusta} to do the sums
and $n(p)$ is the usual Bose function (\ref{Bose factor}).
The result is zero because (1) the temperature-independent piece
in the penultimate line vanishes by dimensional analysis in $4-2\eps$
dimensions, and (2) the two terms $1/(P+Q)^2$ and $1/(P-Q)^2$ in brackets
exactly cancel each other.  Because of this cancelation,
the full result (\ref{Isun result}) for $I_\sun$ is just equal to the
purely three-dimensional contribution (\ref{H333}) at leading order in
the masses.

\subsection {Euclidean derivation}

In other computations in this paper, we will need to know the separate
contributions of various subsets of Euclidean modes $(p_0,q_0)$ to
$I_\sun$.  To get the formulas we need, we shall now rederive the result
for $I_\sun$ using purely Euclidean methods, similar to our derivation
of the basketball integral.  Start with
\begin{equation}
    I_{\rm sun} = A_{\rm sun} + B_{\rm sun}
	+ {T^2 \over (4 \pi)^2}
	\ln {m_1 \over m_1 + m_2 + m_3}
	+ O (m,\eps) \,,
\label{I_sun_exp}
\end{equation}
where $A_{\rm sun}$ and $B_{\rm sun}$ are defined by
\begin{equation}
    A_{\rm sun} \equiv
        \sumint'_P {\Pi (P) \over P^2} \,,
\label{A_sun_def}
\end{equation}
\begin{equation}
    B_{\rm sun} \equiv T \mu^{2 \epsilon} \int
        {d^{3-2\epsilon} p \over (2 \pi)^{3-2\epsilon}}
	{\Pi (p_0{=}0,{\vec p}) \over p^2 + m_1^2}
\label{B_sun_def} \,.
\end{equation}

First consider $A_{\rm sun}$.  As usual, we need to subtract out the UV
divergences:
\begin{equation}
    A_{\rm sun} =
        \sumint'_P {\PiZ (P) \over P^2}
	+ \sumint'_P {2\over P^4} \sumint_Q{1\over Q^2}
        + \sumint'_P {1 \over P^2} \left[
            \PiT (P)
            - {2 \over P^2} \sumint_Q{1\over Q^2}
          \right] \,,
\label{A_sun_splitted}
\end{equation}
where we have used the limiting behavior (\ref{large p limit}) of $\PiT(P)$.
The second term is now convergent in four dimensions.
Exploiting the expression (\ref{Pi0_exp}) for $\Pi^{(0)} (P)$
and the integral (\ref{S_n_result}) enables us to evaluate the
divergent parts of $A_{\rm sun}$ as
\begin{eqnarray}
    \sumint'_P {\PiZ (P) \over P^2}
	&=&  {1\over(4\pi)^2} \left({1\over\eps}+2-\gammaE + O(\eps) \right)
              \sumint {1\over P^2} \left(4\pi\mu^2\over P^2\right)^\eps
\nonumber\\
	&=& {T^2 \over (4 \pi)^2}
	\left [{1 \over 12\epsilon}
	+ {1\over3} \lnmub
	+ {1 \over 3} {\zeta'(-1) \over \zeta(-1)}
	+ {1 \over 2} \right ] + O(\epsilon) \,,
\label{A_sun_d1_result}
\end{eqnarray}
\begin {equation}
     \sumint'_P {2\over P^4} \sumint_Q{1\over Q^2}
	= {T^2 \over (4 \pi)^2}
	\left [{1 \over 6\epsilon}
	+ {2\over3} \lnmub
	+ {1 \over 3} {\zeta'(-1) \over \zeta(-1)}
        + {1 \over 3} \gammaE
	+ {1 \over 3} \right ] + O(\epsilon) \,.
\label{A_sun_d2_result}
\end{equation}
The finite part of $A_{\rm sun}$ is calculated by utilizing
the coordinate space integral representation~(\ref{PiT integral})
for $\Pi^{(T)}(P)$ and (\ref{PiT limit integral}) for its
UV behavior.
Performing the ${\vec p}$ integral and the $p_0$ sum produces
\begin{eqnarray}
   \sumint'_P
	{1 \over P^2} \left [ \PiT (P)
          - {2\over P^2} \sumint_Q{1\over Q^2}
	\right ]
   &=&
   {T^2 \over (4 \pi)^2}
	\int_0^{\infty} d \bar r {1 \over \bar r}
         \left (\coth \bar r - 1 \right )
	 \left ( \coth \bar r - {1 \over \bar r} - {\bar r \over 3} \right )
        + O (\epsilon)
\nonumber\\
   &=&
   {T^2 \over (4 \pi)^2}
	\left[
	- {2 \over 3} {\zeta'(-1) \over \zeta(-1)}
	- {1\over3}\gamma_E + \ln(2\pi) - {1 \over 3} \right ]
	+ O (\epsilon) \,,
\label{A_sun_f_result}
\end{eqnarray}
where the method of appendix~\ref{integration appendix}
have been used for the $\bar r$ integral.
Adding (\ref{A_sun_d1_result}, \ref{A_sun_d2_result}, \ref{A_sun_f_result})
yields
\begin{equation}
    A_{\rm sun}
    = \sumint'{\Pi(P)\over P^2}
    = {T^2 \over (4 \pi)^2}
	\left [{1 \over 4 \epsilon}
	+ \ln {{\bar \mu} \over 4 \pi T}
	+ \ln(2\pi) + {1 \over 2} \right ]
	+ O (\epsilon) \,.
\label{A_sun_result}
\end{equation}

Now consider $B_{\rm sun}$.
Though (\ref{B_sun_def}) contains
ultraviolet divergences due to the zero temperature part of $\Pi (P)$,
dimensional analysis shows that the
contribution from the zero temperature part is $O(m)$ and so
need not concern us at leading order.
Making use of (\ref{PiT integral}) and completing
the ${\vec p}$ integral,
\begin{eqnarray}
    B_{\rm sun} &=& {T^2 \over (4 \pi)^2}
	\int d^3 r {e^{- m_1 r} \over 4 \pi r}
	{1 \over r^2} \left (\coth \bar r - {1 \over \bar r}
	\right ) + O (m, \epsilon)
\nonumber\\
	&=& {T^2 \over (4 \pi)^2}
	\int_0^{\infty} {d \bar r \over \bar r}
	\left [e^{-\bar m_1 \bar r}
	\left (\coth \bar r - {1 \over \bar r} - 1 \right )
	+ e^{-\bar m_1 \bar r} \right ] + O (m, \epsilon)
\nonumber\\
	&=& {T^2 \over (4 \pi)^2}
	\int_0^{\infty} {d \bar r \over \bar r}
        \left [\left (\coth \bar r - {1 \over \bar r} - 1 \right )
        + e^{-\bar m_1 \bar r} \right ] + O (m, \epsilon)
\nonumber\\
	&=& {T^2 \over (4 \pi)^2} \ln {2 T \over m_1}
	+ O (m, \epsilon) \,,
\label{B_sun_result}
\end{eqnarray}
where we have defined $\bar m_1 = m_1/2\pi T$ and done the last step by the
method of Appendix~\ref{integration appendix}.
Combining (\ref{I_sun_exp}, \ref{A_sun_result}, \ref{B_sun_result}) gives
\begin{equation}
    I_{\rm sun}
    = {T^2 \over (4 \pi)^2}
	\left [{1 \over 4 \epsilon}
	+ \ln {{\bar \mu} \over m_1 + m_2 + m_3}
	+ {1 \over 2} \right ] + O (m, \epsilon) \,.
\label{I_sun_result}
\end{equation}

Before leaving this section, we should collect some additional
results that will be
useful elsewhere.\
Subtracting (\ref{H333m1}) from
(\ref{B_sun_result}) gives
\begin {equation}
   \sumint {\delta_{p_0^{}} (1-\delta_{q_0^{}}) \over
           (P^2+m^2) Q^2 (P+Q)^2}
    = {T^2 \over (4 \pi)^2}
	\left [-{1 \over 4 \epsilon}
	+ \ln {2T \over \bar\mu}
	- {1 \over 2} \right ] + O (m, \epsilon) \,.
\label {H344}
\end {equation}
Finally, adding (\ref{A_sun_d2_result}) and (\ref{A_sun_f_result}) gives
\begin{equation}
   \sumint'_P {\PiT(P) \over P^2}
   = {T^2 \over (4 \pi)^2}
	\left [{1 \over 6 \epsilon}
        + {2\over3} \lnmub
	- {1 \over 3} {\zeta'(-1) \over \zeta(-1)}
	+ \ln(2\pi)
        \right ]
	+ O (\epsilon) \,.
\label{sun PiT piece}
\end{equation}


\section {Example of reduction to the scalar basketball}
\label {sample reduction appendix}

Consider the reduction of fig.~\ref{figf}(i):
\begin{equation}
    -\mu^{2\eps}F^{\rm i} = - {1 \over 8} \da \ga^4 \sumint_{PQK}
	{P \cdot (Q-K) \, (P-K) \cdot Q
	\over P^2 Q^2 K^2 (P-Q)^2 (Q-K)^2 (K-P)^2} \,.
\end{equation}
By expanding the numerator as in (\ref{(i) expansion}), we get
\begin{eqnarray}
    -\mu^{2\eps} F^{\rm i}
    =&& {1 \over 16} \da \ga^4 \, \sumint_{PQK} \Biggr [
        - { (P{-}K){\cdot}Q \over P^2 Q^2 K^2 (P{-}Q)^2 (Q{-}K)^2}
	+ {(P{-}K){\cdot}Q \over P^2 Q^2 (P{-}Q)^2 (Q{-}K)^2 (K{-}P)^2}
\nonumber\\
     && + {(P{-}K){\cdot}Q \over P^2 Q^2 K^2 (Q{-}K)^2 (K{-}P)^2}
	- {(P{-}K){\cdot}Q \over P^2 K^2 (P{-}Q)^2 (Q{-}K)^2 (K{-}P)^2}
	\Biggr ] \,.
\end{eqnarray}
Now switch the variables $K$ and $Q$ in the second lines:
\begin{equation}
    -\mu^{2\eps} F^{\rm i} = {1 \over 16} \da \ga^4 \sumint_{PQK}
	\left [
        - {P{\cdot}(Q{-}K) \over P^2 Q^2 K^2 (P{-}Q)^2 (Q{-}K)^2}
	+ {P{\cdot}(Q{-}K) \over P^2 Q^2 (P{-}Q)^2 (Q{-}K)^2 (K{-}P)^2}
        \right ] \,.
\end{equation}
Use the identity~(\ref{trick}) to substitute $P,K{\to}Q/2$ in
the first numerator and $K{\to}(P{+}Q)/2$ in the second:
\begin{eqnarray}
    -\mu^{2\eps} F^{\rm i}
    &=& {1 \over 64} \da \ga^4 \sumint_{PQK}
        \left [
        - {1 \over P^2 K^2 (P{-}Q)^2 (Q{-}K)^2}
	+ {2 P{\cdot}Q{-}2 P^2 \over P^2 Q^2 (P{-}Q)^2 (Q{-}K)^2 (K{-}P)^2}
	\right ]
\nonumber\\
	&=& {1 \over 64} \da \ga^4 \sumint_{PQK}
        \left [
        - {1 \over P^2 K^2 (P{-}Q)^2 (Q{-}K)^2}
	- {1 \over P^2 Q^2 (Q{-}K)^2 (K{-}P)^2} \right ]
\nonumber\\
	&=& - {1 \over 32 } \da \ga^4 \sumint_{PQK}
	{1 \over P^2 Q^2 K^2 (P+Q+K)^2} \,,
\end{eqnarray}
where we have written $2 P{\cdot}Q$ as $P^2{+}Q^2{-}(P{-}Q)^2$ for
the second step and shifted integration variables in the last step.


\section {Derivation of $I_\qcd$}
\label {qcd appendix}

The one-loop self-energy of fig.~\ref{fige} can be reduced, using the
methods discussed after (\ref{reduction example}), to the form
\begin {equation}
   \Pi_{\mu\nu}^{ab}(P) =
   \ga^2 \delta^{ab} \left[
       {d-2\over2} \bar\Pi_{\mu\nu}(P)
       - 2 (P^2\delta_{\mu\nu}-P_\mu P_\nu) \sumint_Q{1\over Q^2(P+Q)^2}
   \right] \,,
\label {Iqcd decomposition}
\end {equation}
where
\begin {equation}
   \bar\Pi_{\mu\nu} \equiv
   2 \delta_{\mu\nu} \sumint_Q {1\over Q^2}
   - \sumint_Q {(2Q+P)_\mu (2Q+P)_\nu \over Q^2 (P+Q)^2} \,.
\label {Pi qed defn}
\end {equation}
$\bar\Pi_{\mu\nu}$ happens to be the form the self-energy would take
in scalar QED.  We find it convenient to introduce $\bar\Pi_{\mu\nu}$
mostly for reasons historic to our original derivation and
because the decomposition (\ref{Iqcd decomposition}) simplifies some
of the algebra of the following calculation.

By again applying the same reduction methods, one may easily verify that
both (\ref{Iqcd decomposition}) and (\ref{Pi qed defn}) share the property
that $P_\mu \Pi_{\mu\nu} = 0$.  In finite temperature non-Abelian gauge
theory, this is a property of the one-loop self energy which does not
persist to higher loops \cite{Vokos}.
We shall use this property in our derivation.

\subsection {Consequences of $P_\mu \Pi_{\mu\nu} = 0$ at one loop}

The orthogonality of $\Pi_{\mu\nu}$ to $P_\mu$ implies that it can be
decomposed into separate transverse and longitudinal pieces:%
\footnote{
   For a review, see refs.~\cite{gpy,Kapusta}.
}
\begin {equation}
   \Pi_{\mu\nu}(P) = \Pi_\t(P) \, {\cal P}_{\t\mu\nu}
                   + \Pi_\l(P) \, {\cal P}_{\l\mu\nu} \,,
\label {lt decomposition}
\end {equation}
where the Euclidean projection operators are given by
\begin {eqnarray}
   {\cal P}_{\t ij} &=& \delta_{ij} - p_i p_j / p^2 \,,
   \qquad
   {\cal P}_{\t00} = {\cal P}_{\t 0i} = {\cal P}_{\t i0} = 0 \,,
\\
   {\cal P}_{\l\mu\nu} &=& \delta_{\mu\nu} - P_\mu P_\nu / P^2
                    - {\cal P}_{\t\mu\nu} \,.
\end {eqnarray}
(\ref{lt decomposition}) then gives
\begin {equation}
   (\Pi_{\mu\nu})^2 = \Pi_\l^2 + (d-2) \Pi_\t^2
   = \left({P^2\over p^2} \Pi_{00}\right)^2
   + {1\over d-2} \left( \Pi_{\mu\mu} - {P^2\over p^2} \Pi_{00}\right)^2 \,.
\label {pi2 decomposition}
\end {equation}

\subsection {Scalar QED}

We now work to evaluate the integral
\begin {equation}
   I_\sqed \equiv \sumint_P {[\Delta\bar\Pi_{\mu\nu}(P)]^2 \over P^4}
\label {Isqed defn}
\end {equation}
and start by separating out the zero-temperature piece of $\bar\Pi_{\mu\nu}$:
\begin {equation}
   I_\sqed =
      \sumint {1\over P^4}
         \left[ \Delta\bar\PiT_{\mu\nu}(P) \right]^2
      + 2 \sumint {1\over P^4} \bar\PiZ(P)
         \Delta\bar\PiT_{\mu\nu}(P)
      + \sumint {1\over P^4} \left[\bar\PiZ(P)\right]^2 \,.
\label {Isqed decomposition}
\end {equation}

\subsubsection {The [finite temperature]$^{\,2}$ piece}

Let's evaluate the $p_0 \not= 0$ part of the sum for the first
integral.  First apply the standard reduction tricks to obtain
\begin {equation}
   \bar\Pi_{\mu\mu}(P) = P^2 \Pi(P) + 2(d-2) \sumint_Q {1\over Q^2} \,,
\label{trace barPi}
\end {equation}
where $\Pi(P)$ is the scalar integral (\ref{pi def}).
Next, we need to isolate the UV divergence of the $P$ integration by
isolating the large $P$ behavior of $\bar\Pi_{\mu\nu}$:
\begin {equation}
   \bar\PiT_{\mu\mu}
     \mathop{\longrightarrow}\limits_{P\to\infty}
     2(d-1) \sumint{1\over Q^2} \,,
   \qquad
   \bar\PiT_{00}
     \mathop{\longrightarrow}\limits_{P\to\infty}
     {2 p^2\over P^2} \sumint{1\over Q^2} \,.
\end {equation}
Specializing to the finite-temperature pieces of $\bar\Pi_{\mu\nu}$,
(\ref{pi2 decomposition}) can be algebraically rewritten as
\begin {eqnarray}
   \left[\bar\PiT_{\mu\nu}\right]^2
      = && {(d-1) \over (d-2)} {P^4\over p^4}
           \left[\hat\PiT_{00}\right]^2
      - {2\over(d-2)} {P^4\over p^2} \hat\PiT_{00} \hat\PiT
      + {1\over(d-2)} P^4 \left[\PiT\right]^2
\nonumber
\\ &&
      + 4 {(d-3)\over(d-2)} P^2 \PiT \sumint {1\over Q^2}
      + 4\left(d-3+{1\over d-2}\right) \left( \sumint{1\over Q^2} \right)^2 \,,
\end {eqnarray}
where we have introduced the UV subtracted
\begin {eqnarray}
   \hat\PiT_{00}(P) &\equiv& \bar\PiT_{00}(P)
         - {2 p^2\over P^2} \sumint{1\over Q^2} \,,
\\
   \hat\PiT(P) &\equiv& \PiT(P) - {2\over P^2} \sumint{1\over Q^2} \,.
\end {eqnarray}
The integral we want is now
\begin {eqnarray}
   \sumint' {1\over P^4}\left[\bar\PiT_{\mu\nu}\right]^2
      = && {(d-1) \over (d-2)} \sumint' {1\over p^4}
           \left[\hat\PiT_{00}\right]^2
      - {2\over(d-2)} \sumint' {1\over p^2} \hat\PiT_{00} \hat\PiT
      + {1\over(d-2)} \sumint' \left[\PiT\right]^2
\nonumber
\\ &&
      + 4 {(d-3)\over(d-2)} \sumint' {1\over P^2}\PiT
           \sumint {1\over Q^2}
      + 4\left(d-3+{1\over d-2}\right) \sumint'{1\over P^4}
           \left( \sumint{1\over Q^2} \right)^2 \,.
\label {pi2 integrals}
\end {eqnarray}
The integrals in the last three terms can be obtained from
(\ref{quad result}, \ref{quart result},
\ref{PiT2 integral}, \ref{sun PiT piece}).
We need to focus on the first two terms, which are convergent and may be
evaluated with $\eps=0$.
{}From the observation that
\begin {equation}
   \partial_r^2 \left( \sum_{q_0} e^{-|q_0|r} e^{-|p_0+q_0|r} \right)
   = \sum_{q_0} (2q_0+p_0)^2 e^{-|q_0|r} e^{-|p_0+q_0|r}
       + {\textstyle{2\over3}} (2\pi T)^2 e^{-|p_0|r} |\bar p_0| (\bar p_0^2-1)
   \,,
\end {equation}
where $\partial_r^2$ means $d^2/dr^2$ and not $\nabla^2$,
one may easily relate $\bar\Pi_{00}$  to the scalar case (\ref{careless pi}):
\begin {eqnarray}
   \bar\Pi_{00}(P)
   &=& 2\sumint_Q {1\over Q^2}
       - \sumint_Q {(2q_0+p_0)^2 \over Q^2 (P+Q)^2}
\nonumber\\
   &=& {T^2\over6} -
      {T^3\over4} \int d^3r\, {1\over r^2} e^{i \vec p\cdot \vec r}
      \left\{ \partial_{\bar r}^2
         \left[ e^{-|p_0|r} (\coth\bar r + |\bar p_0|) \right]
         - {\textstyle{2\over3}} e^{-|p_0|r} |\bar p_0| (\bar p_0^2-1)
      \right\}
\nonumber\\ && \qquad
   + O(\eps) \,,
\\
   \hat\PiT_{00}(P)
   &=& - {T^3\over4} \int d^3r\, {1\over r^2} e^{i \vec p\cdot \vec r}
       \partial_{\bar r}^2 \left[ e^{-|p_0|r}
          \left(\coth\bar r - {1\over\bar r} - {\bar r\over 3}\right) \right]
   + O(\eps) \,.
\end {eqnarray}
Performing the angular integration and then integrating by parts yields
\begin {eqnarray}
   \hat\PiT_{00}(P)
   &=& -{T\over4\pi} \int\nolimits_0^\infty \! dr \,
       \left( \partial_r^2 {\sin pr \over pr} \right) e^{-|p_0|r}
       \left(\coth\bar r - {1\over\bar r} - {\bar r\over 3}\right)
   + O(\eps) \,.
\end {eqnarray}
The first two integrals in (\ref{pi2 integrals}) are then
\begin {eqnarray}
   {3\over2} \sumint' {1\over p^4}
      \left[\hat\PiT_{00}\right]^2
   &=& {3\over2} {T^3\over(4\pi)^2} \sum_{p_0 \not= 0}
      \int\nolimits_0^\infty \! dr \int\nolimits_0^\infty \! ds \,
      e^{-|p_0|(r+s)}
      \left(\coth\bar r - {1\over\bar r} - {\bar r\over3}\right)
      \left(\coth\bar s - {1\over\bar s} - {\bar s\over3}\right)
\nonumber
\\ && \qquad \times
      {1\over(2\pi)^3} \int {d^3p\over p^4}
      \partial_r^2 {\sin pr\over pr} \partial_s^2 {\sin ps\over ps}
   ~~+~~ O(\eps) \,,
\label {eq Abar}
\\
   - \sumint' {1\over p^2} \hat\PiT_{00} \hat\PiT
   &=& {T^3\over(4\pi)^2} \sum_{p_0 \not= 0}
      \int\nolimits_0^\infty \! dr \int\nolimits_0^\infty \! ds \,
      e^{-|p_0|(r+s)}
      \left(\coth\bar r - {1\over\bar r} - {\bar r\over3}\right)
      \left(\coth\bar s - {1\over\bar s} - {\bar s\over3}\right)
\nonumber
\\ && \qquad \times
      {1\over(2\pi)^3} \int {d^3p\over p^2}
      \partial_r^2 {\sin pr\over pr} {\sin ps\over ps}
   ~~+~~ O(\eps) \,.
\label {eq Bbar}
\end {eqnarray}
Now plug in
\begin {eqnarray}
   {1\over(2\pi)^3} \int {d^3p\over p^4}
      \partial_r^2 {\sin pr\over pr} \partial_s^2 {\sin ps\over ps}
   &=& -{1\over6\pi r_>^3} + {1\over4\pi r^2} \delta(r-s) \,,
\\
   {1\over(2\pi)^3} \int {d^3p\over p^2}
      \partial_r^2 {\sin pr\over pr} {\sin ps\over ps}
   &=& {1\over2\pi r^3} \theta(r-s) - {1\over4\pi r^2} \delta(r-s) \,.
\end {eqnarray}
Amazingly, the terms not proportional to $\delta(r-s)$ cancel in the
sum of (\ref{eq Abar}) and (\ref{eq Bbar}).  The $\delta(r-s)$ terms then
give:
\begin {equation}
   {3\over2} \sumint' {1\over p^4}
      \left[\hat\PiT_{00}\right]^2
   - \sumint' {1\over p^2} \hat\PiT_{00} \hat\PiT
   = {1\over2} \sumint' \left[ \hat\PiT \right]^2
   + O(\eps) \,.
\label{cancellation result}
\end {equation}
This could be easily evaluated using the techniques of
appendix~\ref{integration appendix}, but we'll leave it in this form for now.

The evaluation of the $p_0{=}0$ piece of (\ref{Isqed decomposition})
proceeds in much the same way, but we don't need to make any UV
subtractions.  One finds
\begin {eqnarray}
   \Delta\bar\PiT_{00}(0,p)
   &=& - {T^3\over4} \int d^3r\, {1\over r^2}
       \left( e^{i \vec p\cdot \vec r} - 1 \right)
       \partial_{\bar r}^2
          \left(\coth\bar r - {1\over\bar r}\right)
   + O(\eps)
\nonumber\\
   &=& - {T\over4\pi} \int\nolimits_0^\infty dr
       \left( \partial_r^2 {\sin pr \over pr} \right)
          \left(\coth\bar r - {1\over\bar r}\right)
   + O(\eps) \,.
\end {eqnarray}
The integral is
\begin {eqnarray}
   T\int {d^3p \over (2\pi)^3} {1\over p^4}
      \left[\Delta\bar\PiT_{\mu\nu}(0,p)\right]^2
   &=& T\int {d^3p \over (2\pi)^3} \left\{
      {3\over2} {1\over p^4} \left[\Delta\bar\PiT_{00}\right]^2
      - {1\over p^2} \Delta\bar\PiT_{00} \PiT
      + {1\over2} \left[\PiT\right]^2
   \right\}
\nonumber\\ && \qquad
   + O(\eps) \,.
\end {eqnarray}
The same sort of cancelation occurs between the first two terms as in
the $p_0 \not= 0$ case, and we are left with
\begin {equation}
   \sumint {\delta_{p_0^{}} \over P^4}
      \left[\Delta\bar\PiT_{\mu\nu}\right]^2
   = \sumint \delta_{p_0^{}} \left[ \PiT \right]^2
   + O(\eps) \,.
\end {equation}
Putting this together with the $p_0 \not= 0$ results
(\ref{pi2 integrals}) and (\ref{cancellation result})
yields, after a little reorganization,
\begin {equation}
   \sumint {1\over P^4} \left[\Delta\bar\PiT_{\mu\nu}\right]^2
   = \sumint \left[ \PiT \right]^2
   + 4(d-2) \sumint{1\over P^4} \left( \sumint{1\over Q^2}\right)^2
   + O(\eps) \,.
\label {sqed TT result}
\end {equation}
The first integral is given by (\ref{PiT2 integral}).
One wonders if there's an easier way to get (\ref{sqed TT result}).

\subsubsection {The rest of it}

The cross-term between $\bar\PiT_{\mu\nu}$ and $\bar\PiZ_{\mu\nu}$ is
easy because $\bar\PiZ_{\mu\nu}$ is proportional to
$P^2\delta_{\mu\nu} - P_\mu P_\nu$.
Using (\ref{trace barPi}),
\begin {eqnarray}
   \sumint {1\over P^4} \bar\PiT_{\mu\nu} \bar\PiZ_{\mu\nu}
   &=& {1\over d-1} \sumint {1\over P^2} \bar\PiT_{\mu\mu} \PiZ
\nonumber\\
   &=& {1\over d-1} \sumint \PiT \PiZ
      + 2 {(d-2)\over(d-1)} \sumint {1\over P^2} \PiZ \, \sumint{1\over Q^2}
   \,.
\label {sqed T0 result}
\end {eqnarray}
The integrals can be found in
(\ref{quad result}, \ref{secondterm}, \ref{A_sun_d1_result}).  The final
integral we need is
\begin {equation}
   \sumint {1\over P^4} \left[ \bar\PiZ_{\mu\nu} \right]^2
   = {1\over d-1} \sumint \left[ \PiZ \right]^2 \,,
\label {sqed 00 result}
\end {equation}
which may be found in (\ref{thirdterm}).
Combining (\ref{Isqed decomposition}, \ref{sqed TT result},
\ref{sqed T0 result}, \ref{sqed 00 result}), and incorporating the results
for the assorted basic integrals, gives
\begin {equation}
   I_\sqed
   = {1\over(4\pi)^2} \left(T^2\over12\right)^2 \left[
        {46\over3\eps} + 92 \lnmub
        - {44\over3} {\zeta'(-3)\over\zeta(-3)}
        + {272\over3} {\zeta'(-1)\over\zeta(-1)}
        + 16 \gammaE
        + {1034\over15} \right]
   + O(\eps) \,.
\label {Isqed result}
\end {equation}

\subsection {Non-abelian gauge theory}

Using (\ref{Iqcd decomposition}) and our standard reduction tricks, it is
easy to obtain
\begin {eqnarray}
   I_\qcd =
      \left(d-2 \over 2\right)^2 I_\sqed
      + 2d I_\bball
      - 4 (d-2)^2 \sumint' {1\over P^2} \Pi(P) \sumint {1\over Q^2}
   \,,
\label {Iqcd result}
\end {eqnarray}
which, when combined with
(\ref{quad result}, \ref{A_sun_result}, \ref{Isqed result}),
is our final result for $I_\qcd$.


\end {document}